%% file: LNAI_GP_revised2arxiv.tex
\documentclass[runningheads,a4paper,12pt]{llncs}

\usepackage{amsfonts}
\usepackage{amssymb}
\usepackage{amsbsy} % Per a \boldsymbol
\usepackage{amsmath}
\usepackage{subfigure}
\usepackage[dvips]{graphicx}
\usepackage[T1]{fontenc}
\usepackage[utf8]{inputenc}
\usepackage{color}
\usepackage{array}
\usepackage{animate}
\usepackage[margin=2.8cm]{geometry}

\newcommand{\blue}[1]{\textcolor[rgb]{0,0,0.5}{#1}}

\usepackage{hyperref}
\hypersetup{
    bookmarks=true,         % show bookmarks bar?
    unicode=false,          % non-Latin characters in Acrobat’s bookmarks
    pdftoolbar=true,        % show Acrobat’s toolbar?
    pdfmenubar=true,        % show Acrobat’s menu?
    pdffitwindow=false,     % window fit to page when opened
    pdfstartview={FitH},    % fits the width of the page to the window
    pdftitle={LNAI GP},    % title
    pdfauthor={Camps-Valls},     % author
    pdfsubject={LNAI GP},   % subject of the document
    pdfcreator={Camps-Valls},   % creator of the document
    pdfproducer={Camps-Valls}, % producer of the document
    pdfkeywords={keyword1} {key2} {key3}, % list of keywords
    pdfnewwindow=true,      % links in new window
    colorlinks=true,       % false: boxed links; true: colored links
    linkcolor=blue,          % color of internal links (change box color with linkbordercolor)
    citecolor=blue,        % color of links to bibliography
    filecolor=blue,      % color of file links
    urlcolor=blue           % color of external links
}

\newcommand{\IG}{\includegraphics}
\usepackage{multicol}
\usepackage{float}
\usepackage{lscape}
\usepackage{cite}

\input{notation}

\usepackage[usenames,dvipsnames]{xcolor}
\definecolor{mycolor}{rgb}{0.1, 0.5, 0.2}

\usepackage{url}

\urldef{\mailsa}\path|{fmateo, jordi, lapeva, jverrelst, gcamps}@uv.es|  
\newcommand{\keywords}[1]{\par\addvspace\baselineskip
\noindent\keywordname\enspace\ignorespaces#1}

\begin{document}

\mainmatter  % start of an individual contribution

% first the title is needed
\title{Learning Structures in Earth Observation Data with Gaussian Processes}

% a short form should be given in case it is too long for the running head
\titlerunning{Gaussian Processes in Earth Observation}

% the name(s) of the author(s) follow(s) next
%
% NB: Chinese authors should write their first names(s) in front of
% their surnames. This ensures that the names appear correctly in
% the running heads and the author index.
%
\author{Fernando Mateo \and Jordi Muñoz-Marí \and Valero Laparra \and\\ Jochem Verrelst \and Gustau Camps-Valls
\thanks{\bf Paper published in Advanced Analysis and Learning on Temporal Data. AALTD 2015. Lecture Notes in Computer Science, vol 9785. Springer, Cham. \url{https://doi.org/10.1007/978-3-319-44412-3_6}}}

\authorrunning{Mateo et al.}
% (feature abused for this document to repeat the title also on left hand pages)

% the affiliations are given next; don't give your e-mail address
% unless you accept that it will be published
\institute{Image Processing Laboratory, University of Valencia,\\
C/ Catedrático José Beltrán 2, 46980 Paterna, Spain\\
\mailsa\\
\url{http://isp.uv.es/}}

%
% NB: a more complex sample for affiliations and the mapping to the
% corresponding authors can be found in the file "llncs.dem"
% (search for the string "\mainmatter" where a contribution starts).
% "llncs.dem" accompanies the document class "llncs.cls".
%

%\toctitle{Lecture Notes in Computer Science}
%\tocauthor{Authors' Instructions}
%\ackname{kkk}

\maketitle

\begin{abstract}
Gaussian Processes (GPs) has experienced tremendous success in geoscience in general and for bio-geophysical parameter retrieval in the last years. GPs constitute a solid Bayesian framework to formulate many function approximation problems
consistently. This paper reviews the main theoretical GP developments in the field. We review new algorithms that respect the signal and noise characteristics, that provide feature rankings automatically, and that allow applicability of associated uncertainty intervals to transport GP models in space and time. All these developments are illustrated in the field of geoscience and remote sensing at a local and global scales through a set of illustrative examples. 
\end{abstract}

\section{Introduction}\label{sec:intro}

Spatio-temporally explicit, quantitative retrieval methods of Earth surface and atmosphere characteristics are a requirement in a variety of Earth observation applications. Optical sensors mounted on-board Earth observation (EO) satellites are being endowed with high temporal, spectral and spatial resolutions, and thus enable the retrieval and monitoring of climate and bio-geophysical variables~\cite{Dorigo2007,schaepman09}. With the super-spectral Copernicus Sentinel-2 (S2)~\cite{Drusch2012} and the forthcoming Sentinel-3 missions~\cite{Donlon12}, among other planned space missions, an unprecedented data stream for land, ocean and atmosphere monitoring will soon become available to a diverse user community. This vast data streams require enhanced processing techniques. Statistical inference methods play an important role in this area of research. Understanding is more challenging than predicting, and thus statistical models should not only be accurate but also capture plausible physical relations and explain the problem at hand. 

Over the last few decades a wide diversity of bio-geophysical retrieval methods have been developed, but only a few of them made it into operational processing chains. Essentially, we may find two main approaches to the inverse problem of estimating biophysical parameters from spectra: {\em parametric physically-based models} and {\em non-parametric statistical models}.
Lately, machine learning has attained outstanding results in the estimation of climate variables and related bio-geophysical parameters at local and global scales~\cite{CampsValls11mc}. For example, current operational vegetation products, like leaf area index (LAI), are typically produced with neural networks~\cite{Baret2013}, Gross Primary Production (GPP) as the largest global CO$_2$ flux driving several ecosystem functions is estimated using ensembles of random forests and neural networks~\cite{Beer10,Jung11}, biomass has been estimated with stepwise multiple regression~\cite{Sarker11}, PCA and piecewise linear regression for sun-induced fluorescence (SIF) estimation~\cite{Guanter14}, support vector regression showed high efficiency in modelling LAI, fractional vegetation cover (fCOVER), evapotranspiration~\cite{Yang06,Durbha07}, relevance vector machines were successful in ocean chlorophyll estimation~\cite{Camps-Valls2006}, and recently, Gaussian Processes (GPs)~\cite{Rasmussen06}  provided excellent results in vegetation properties estimation~\cite{Verrelst12rse,Verrelst12b,Verrelst2013a,Roelofsen2014}. 
The family of Bayesian non-parametrics, and of Gaussian processes in particular~\cite{Rasmussen06}, have been payed wide attention in the last years in remote sensing data analysis. We will review the main developments in GPs for EO data analysis in this paper.

The remainder of the paper is organized in two main parts: Section II reviews the main notation and theory of GP regression. Section III presents some of the most recent advances of GP models applied to remote sensing data processing. Section IV presents ways to extract knowledge from those GP models. We conclude in Section V with a discussion about the upcoming challenges and research directions.

\section{Gaussian Process Regression}\label{sec:GPR}

\subsection{Gaussian processes: a gentle introduction}

Gaussian processes (GPs) are state-of-the-art tools for discriminative machine learning. They can be interpreted as a family of kernel methods with the additional advantage of providing a full conditional statistical description for the predicted variable. Standard regression approximates observations (often referred to as \emph{outputs}) $\{y_n\}_{n=1}^{N}$  as the sum of some unknown latent function $f(\x)$ of the inputs $\{\x_n \in\Real^D \}_{n=1}^{N}$ plus \emph{constant power (homoscedastic)} Gaussian noise, i.e. 
\begin{equation}\label{GLR}
y_n = f(\x_n) + \varepsilon_n,~~~\varepsilon_n \sim\Normal(0,\sigma^2).
\end{equation}
GP regression proceeds in a Bayesian, non-parametric way, to fit the observed data. A zero mean\footnote{It is customary to subtract the sample mean to data $\{y_n\}_{n=1}^N$, and then to assume a zero mean model.} GP prior is placed on the latent function $f(\vect{x})$ and a Gaussian prior is used for each latent noise term $\varepsilon_n$, 
$f(\vect{x})\;\sim\;\GP(\vect{0}, k_\vect{\theta}(\vect{x},\vect{x}'))$, 
where $k_\vect{\theta}(\vect{x},\vect{x}')$ is a covariance function parametrized by $\vect{\theta}$ and $\sigma^2$ is a hyperparameter that specifies the noise power.
Essentially, a GP is a stochastic process whose marginals are distributed as a multivariate Gaussian. In particular, given the priors $\GP$, samples drawn from $f(\x)$ at the set of locations $\{\x_n\}_{n=1}^N$ follow a joint multivariate Gaussian with zero mean and covariance (sometimes referred as to {\em kernel}) matrix $\mat{K_\vect{ff}}$ with  $[\mat{K_\vect{ff}}]_{ij} = k_\vect{\theta}(\vect{x}_i,\vect{x}_j)$.

If we consider a test location $\x_*$ with corresponding output $y_*$, priors $\GP$ induce a prior distribution between the observations $\y \equiv \{y_n\}_{n=1}^N$ and $y_*$.
\iffalse
\begin{equation*}
  [\!\! 
    \begin{array}{c}
      \vect{y} \\
      y_*
    \end{array}
    \!\!]
  \;\sim\;
  \Normal( \vect{0},\;[\!\!
      \begin{array}{cc}
        \mat{K}_{\vect{ff}}+\sigma^2\mat{I}_n & \vect{k}_{\vect{f}*}\\
        \vect{k}_{\vect{f}*}^\top & k_{**}+\sigma^2\\
      \end{array}
      \!\!]).
      \label{eq:jointprior}
\end{equation*}
\fi
Collecting available data in $\dataset\equiv\{\vect{x}_n,y_n|n=1,\ldots
N\}$, it is possible to analytically compute the posterior distribution over the unknown output $y_*$: 
\begin{align}
 \prob(y_*|\vect{x}_*,\dataset)&=\Normal(y_*|\mu_{\text{GP}*},\sigma_{\text{GP}*}^2) \\
\mu_{\text{GP}*} &= \vect{k}_{\vect{f}*}^\top (\mat{K}_{\vect{ff}}+\sigma^2\mat{I}_n)^{-1}\vect{
y} = \vect{k}_{\vect{f}*}^\top\boldsymbol{\alpha} \label{eq:mugp}\\
\sigma_{\text{GP}*}^2 &= \sigma^2+k_{**}-
     \vect{k}_{\vect{f}*}^\top (\mat{K}_{\vect{ff}}+\sigma^2\mat{I}_n)^{-1}\vect{k}_{\vect
{f}*}. \label{eq:sigmagp}
\end{align}
which is computable in $\bigO(n^3)$ time (this cost arises from the inversion of the $n\times n$ matrix ($\mat{K}_{\vect{ff}}+\sigma^2\mat{I}$), see \cite{Rasmussen06}. %, Ch.~8.
In addition to the computational cost, GPs require large memory since in naive implementations one has to store the training kernel matrix, which amounts to ${\mathcal O}(n^2)$.

\subsection{On the model selection}

The corresponding hyperparameters $\{\vect{\theta},\sigma_n\}$ are typically selected by Type-II Maximum Likelihood, using the marginal likelihood (also called evidence) of the observations, which is also analytical (explicitly conditioning on $\vect{\theta}$ and $\sigma_n$):
\begin{equation}
\log p(\vect{y}|\vect{\theta},\sigma_n) =  \log \Normal(\y|\vect{0}, \Kf + \sigma_n^2\mat{I}).
\label{eq:logevidence}
\end{equation}
When the derivatives of \eqref{eq:logevidence} are also analytical, which is often the case, conjugated gradient ascend is typically used for optimization. 

\subsection{On the covariance function}

The core of a kernel method like GPs is the appropriate definition of the covariance (or kernel) function. A standard, widely used covariance function is the squared exponential,  $k(\vect{x}_i,\vect{x}_j) = \exp(-\|\vect{x}_i-\vect{x}_j\|^2/(2\sigma^2))$, which captures sample similarity well in most of the (unstructured) problems, and only one hyperparameter $\sigma$ needs to be tuned.

\begin{table}[t!]
\begin{center}
\caption{\label{kernelitos}Some kernel functions used in the literature.}

\renewcommand{\tabcolsep}{4pt}
\begin{tabular}{|l|l|}
\hline
{\bf Kernel function} & {\bf Expression}\\
\hline
\hline
Linear                                    & ${k({\bf x},{\bf x}') = {\bf x}^\top {\bf x}' + c}$  \\
Polynomial                                & ${k({\bf x},{\bf x}') = (\alpha {\bf x}^\top {\bf x}' + c)^d}$  \\
Gaussian                                  & ${k({\bf x},{\bf x}') = \exp(-\|{\bf x} - {\bf x}'\|^2 /(2\sigma^2))}$  \\
Exponential                                & ${k({\bf x},{\bf x}') = \exp(-\|{\bf x} - {\bf x}'\| /(2\sigma^2))}$  \\
Rational Quadratic                         & ${k({\bf x},{\bf x}') = 1 - (\|{\bf x} - {\bf x}'\|^2)/(\|{\bf x} - {\bf x}'\|^2 + c)}$  \\
Multiquadric                               & ${k({\bf x},{\bf x}') = \sqrt{\|{\bf x} - {\bf x}'\|^2 + c^2}}$  \\
Inv. Multiquad.                       & ${k({\bf x},{\bf x}') = 1/(\sqrt{\|{\bf x} - {\bf x}'\|^2 + \theta^2})}$    \\
Power                                      & ${k({\bf x},{\bf x}') = - \|{\bf x} - {\bf x}'\|^d}$  \\
Log                                        & ${k({\bf x},{\bf x}') = - \log (\|{\bf x} - {\bf x}'\|^d + 1)}$  \\
\hline
\end{tabular}
\end{center}
\end{table}

In the context of GPs, kernels with more hyperparameters can be efficiently inferred. This is an  opportunity to exploit asymmetries in the feature space by including a parameter per feature, as in the very common anisotropic squared exponential (SE) kernel function:
$$k(\x_i,\x_j) = \nu \exp\bigg(-\sum_{f=1}^F \dfrac{(x_i^f-x_j^f)^2}{2\sigma_f^2}\bigg) + \sigma_n^2\delta_{ij},$$
where $\nu$ is a scaling factor, $\sigma_n$ is the standard deviation of the (estimated) noise, and a $\sigma_f$ is the length-scale per input features, $f=1,\ldots,F$.  This is a very flexible covariance function that typically suffices to tackle most of the problems. Table~\ref{kernelitos} summarizes the most common kernel functions in standard applications with kernel methods.

\subsection{Gaussian processes exemplified}

Let us illustrate the solution of GP regression (GPR) in a toy example. In Fig.~\ref{Fig1ExampleGaussian} we include an illustrative example with 6 training points in the range between $-2$ and $+2$. We firstly depict several random functions drawn from the GP prior and then we include functions drawn from the posterior. We have chosen an isotropic Gaussian kernel and $\sigma_\nu=0.1$. We have plotted the mean function plus/minus two standard deviations (corresponding to a 95\% confidence interval).
Typically, the hyperparameters are unknown, as well as the mean, covariance and likelihood functions. We assumed a Squared Exponential (SE) covariance function and learned the optimal hyperparameters by minimizing the negative log marginal likelihood (NLML) w.r.t. the hyperparameters. 
We observe three different regions in the figure. Below $x=-1.5$, we do not have samples and the GPR provides the solution given by the prior  (zero mean and $\pm 2$). At the center, where most of the data points lie, we have a very accurate view of the latent function with small error bars (close to $\pm 2\sigma_\nu$).  For $x>0$, we do not have training samples neither so we have same behaviour. 
GPs typically provide an accurate solution where the data lies and high error bars where we do not have available information and, consequently, we presume that the prediction in that area is not accurate. This is why in regions of the input space without points the confidence intervals are wide resembling the prior distribution.

\begin{figure}[t!]

\centerline{
\animategraphics[autoplay,loop,width=7.4cm]{5}{./figures/priors_}{01}{20}
\animategraphics[autoplay,loop,width=7.4cm]{5}{./figures/posteriors_}{01}{20}
}

\caption{Example of a Gaussian process. Left: some functions drawn at random from the GP prior. Right: some random functions drawn from the posterior, i.e. the prior conditioned on 6 noise-free observations indicated in big black dots. The shaded area represents the point-wise mean plus and minus two times the standard deviation for each input value (corresponding to the 95 confidence region). It can be noted that the confidence intervals become large for regions far from the observations. \blue{Note: This is an animated figure that only works in Acrobat reader.} }
\label{Fig1ExampleGaussian}
\end{figure}

\section{Advances in Gaussian Process Regression}\label{sec:advances}

In this section, we review some recent advances in GPR especially suited for remote sensing data analysis. We will review the main aspects to design covariance functions that capture non-stationarities and multiscale time relations in EO data, as well as GPs that can learn arbitrary transformations of the observed variable and noise models. 

\subsection{Structured, non-stationary and multiscale}

\begin{figure}[h!]
\begin{center}
\renewcommand{\tabcolsep}{3pt}

\begin{tabular}{cccccc}
{Linear} &{SE} &{Rat. quadratic} &{Periodic} & {Lin+Per}  & {Lin+SE} \\[2mm]
\includegraphics[width=1.8cm,height=1.3cm]{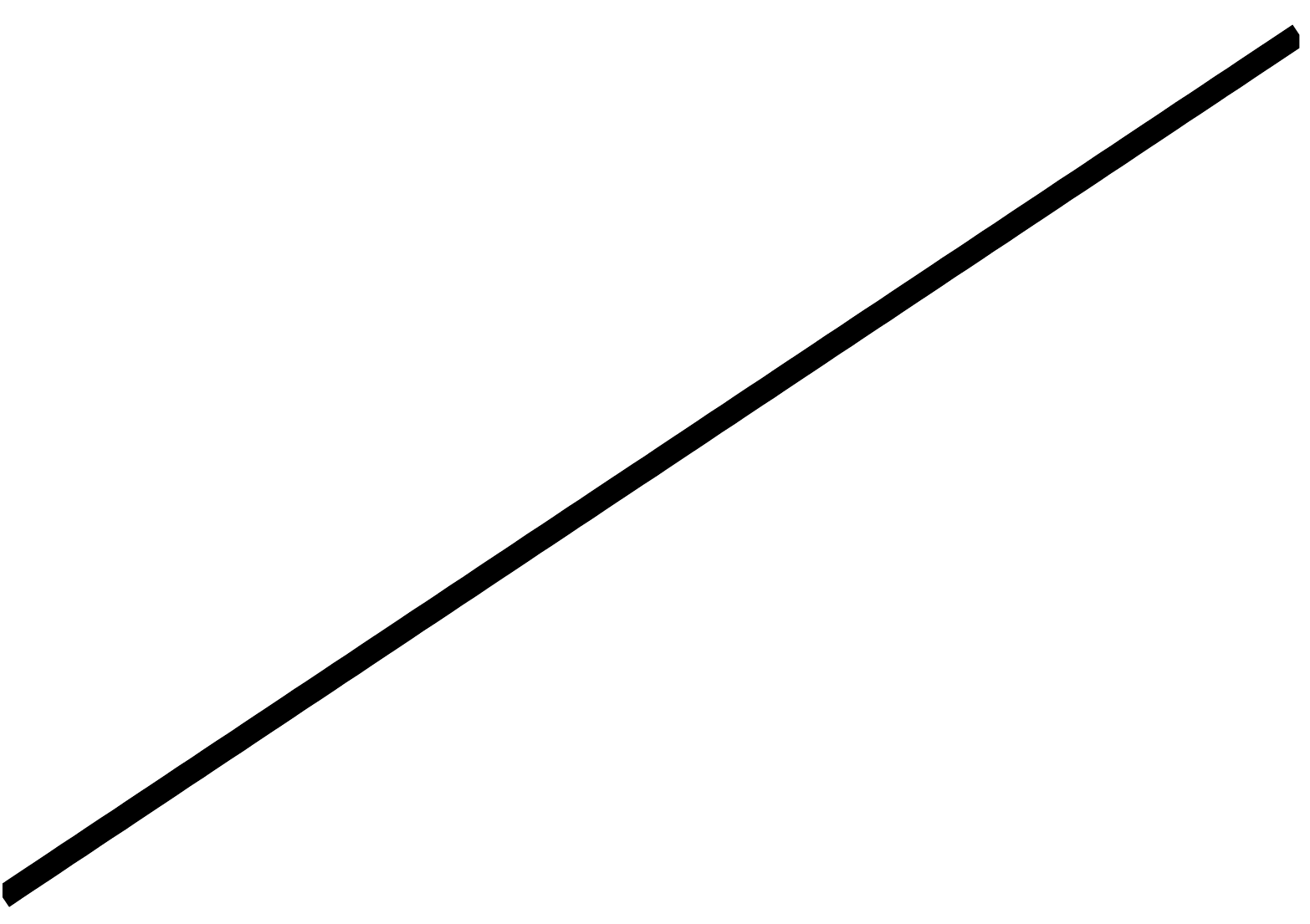} & 
\includegraphics[width=1.8cm,height=1.3cm]{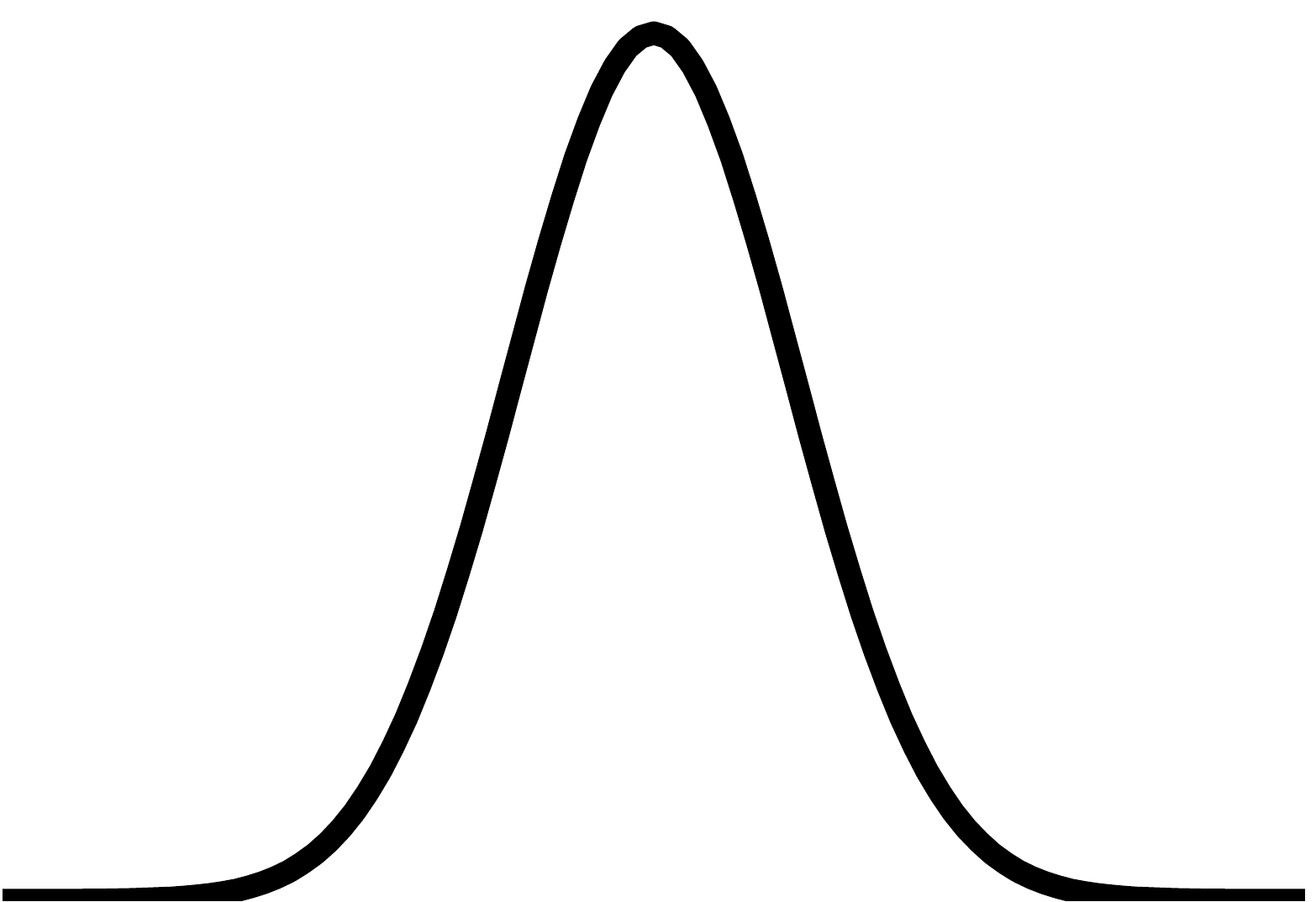} &
\includegraphics[width=1.8cm,height=1.3cm]{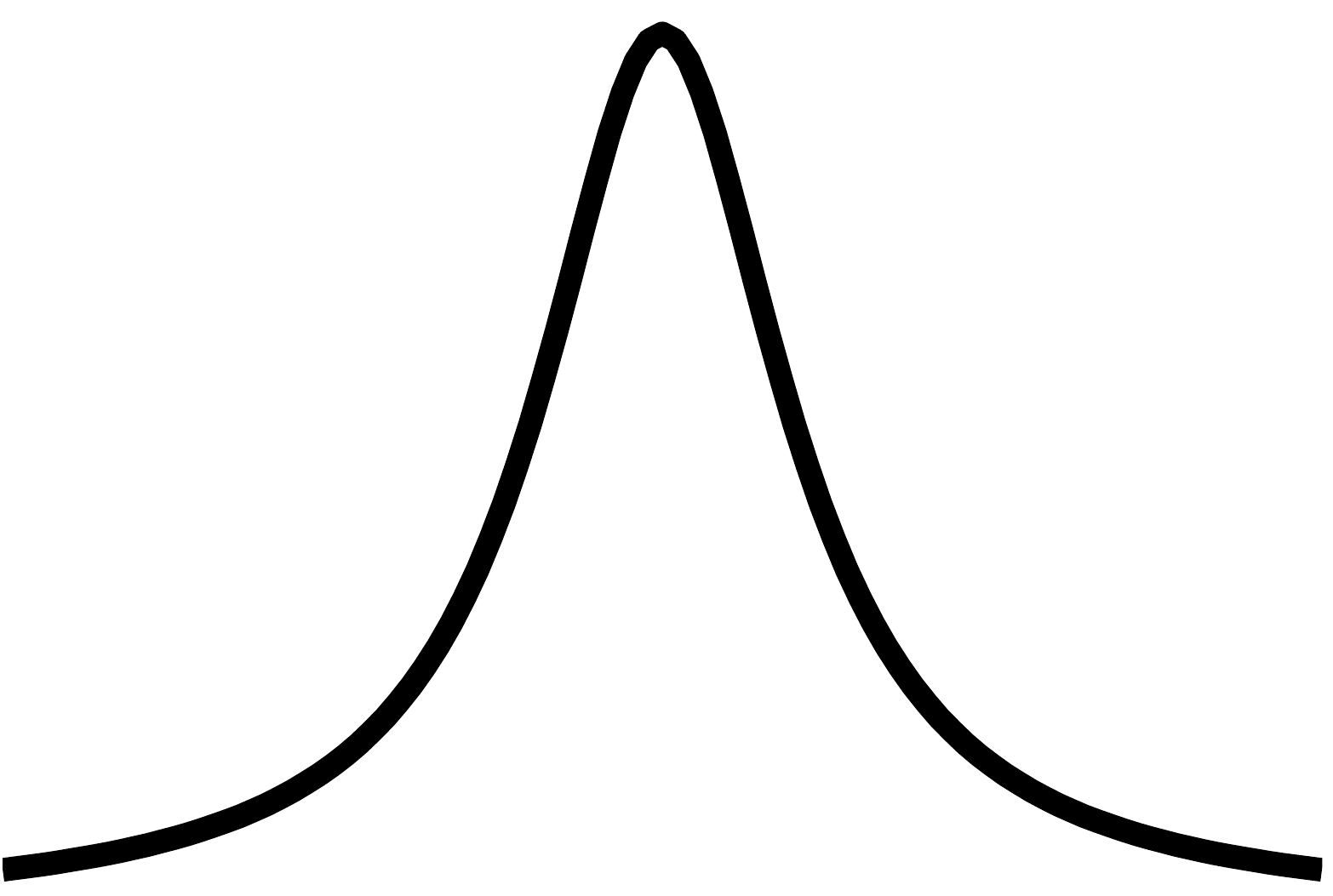} & 
\includegraphics[width=1.8cm,height=1.3cm]{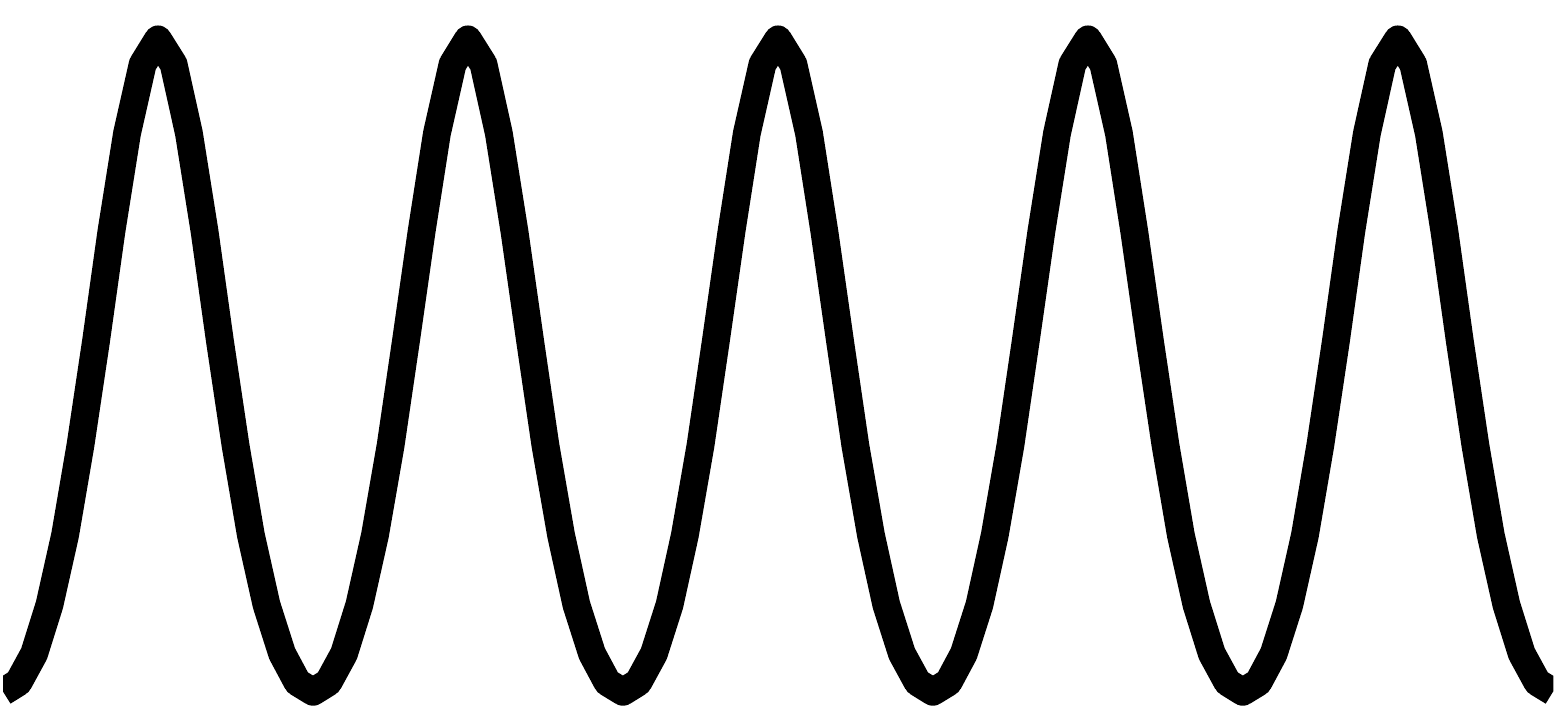} &
\includegraphics[width=1.8cm,height=1.3cm]{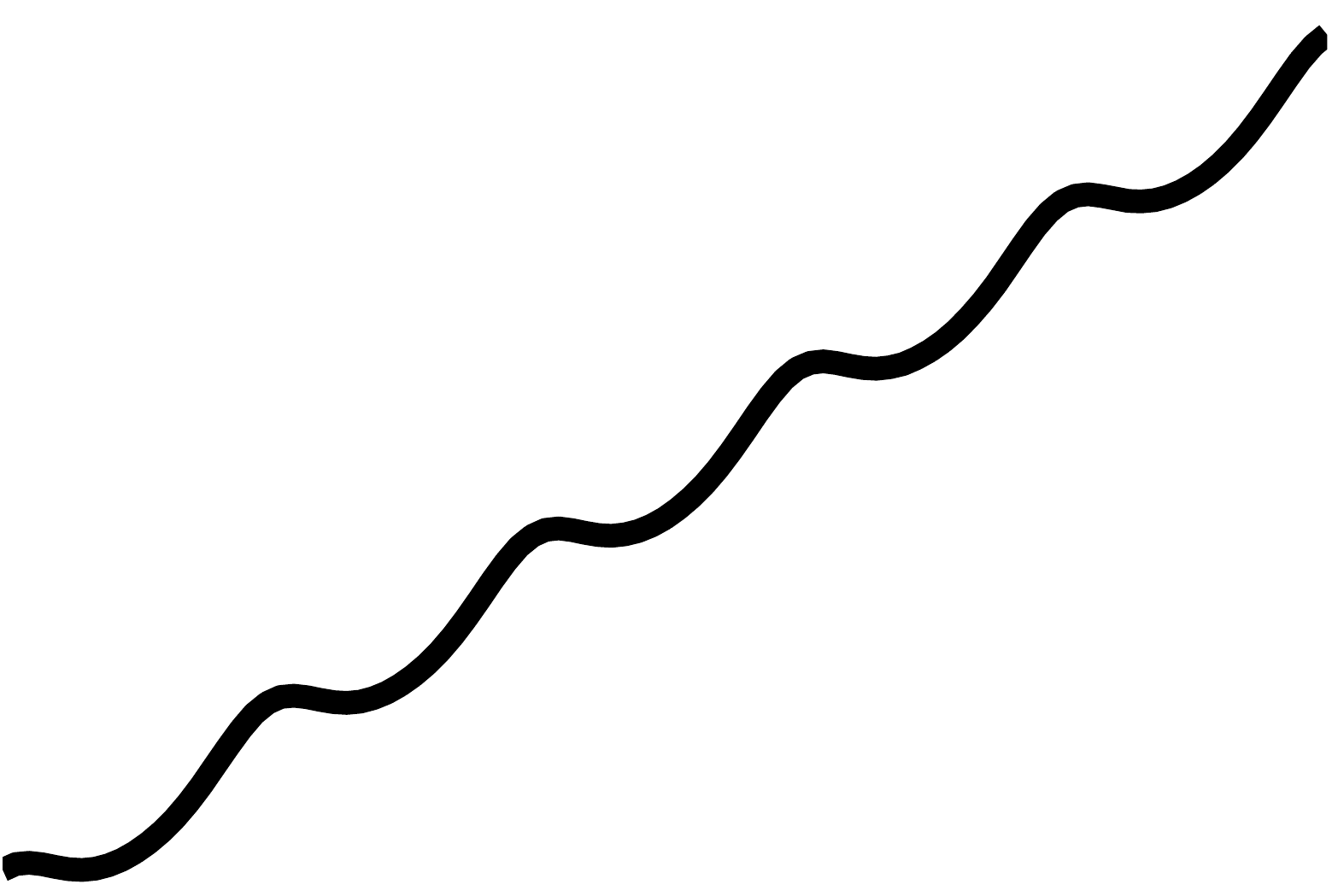} &
\includegraphics[width=1.8cm,height=1.3cm]{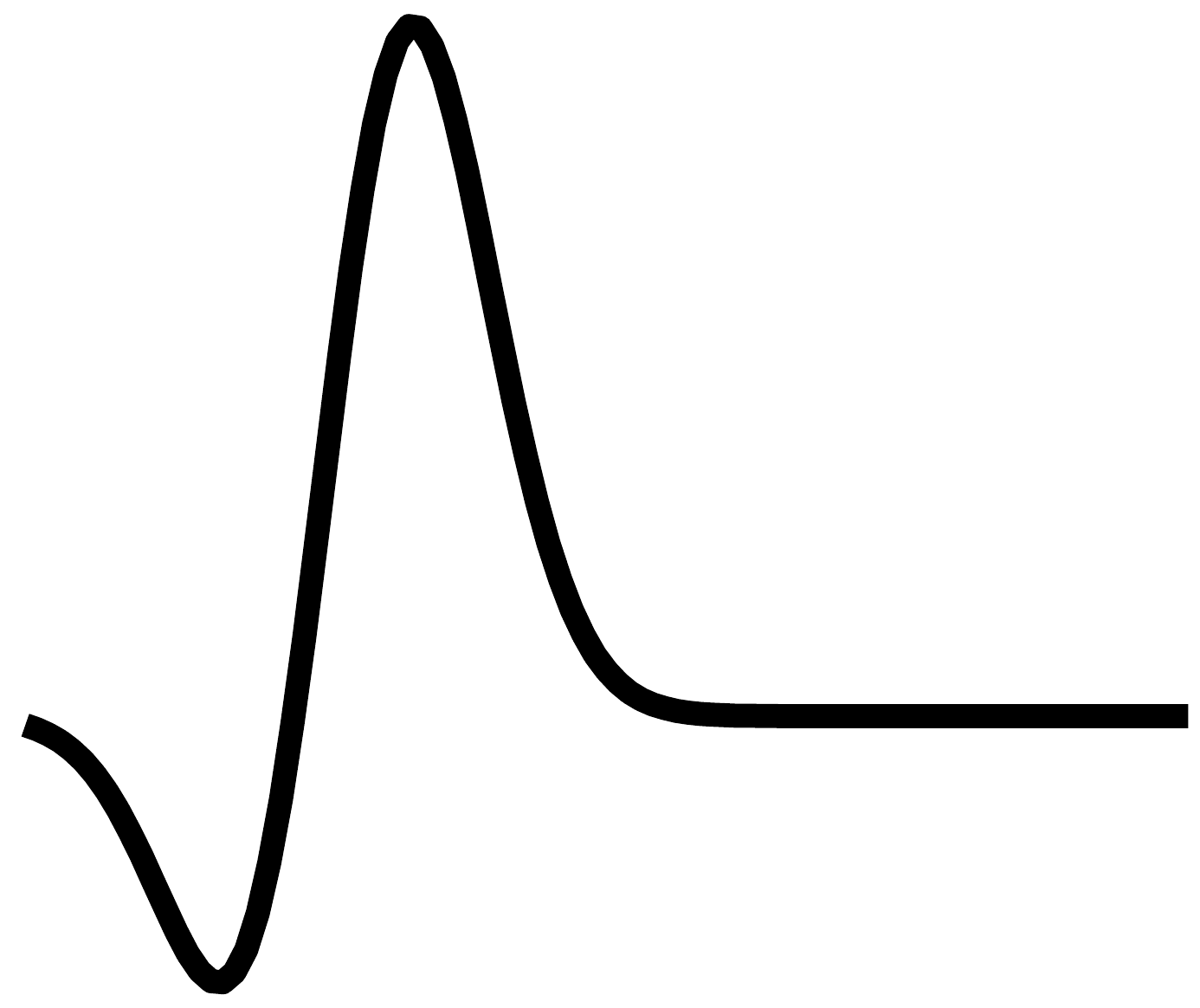} \\[2mm]
\includegraphics[width=1.8cm,height=1.3cm]{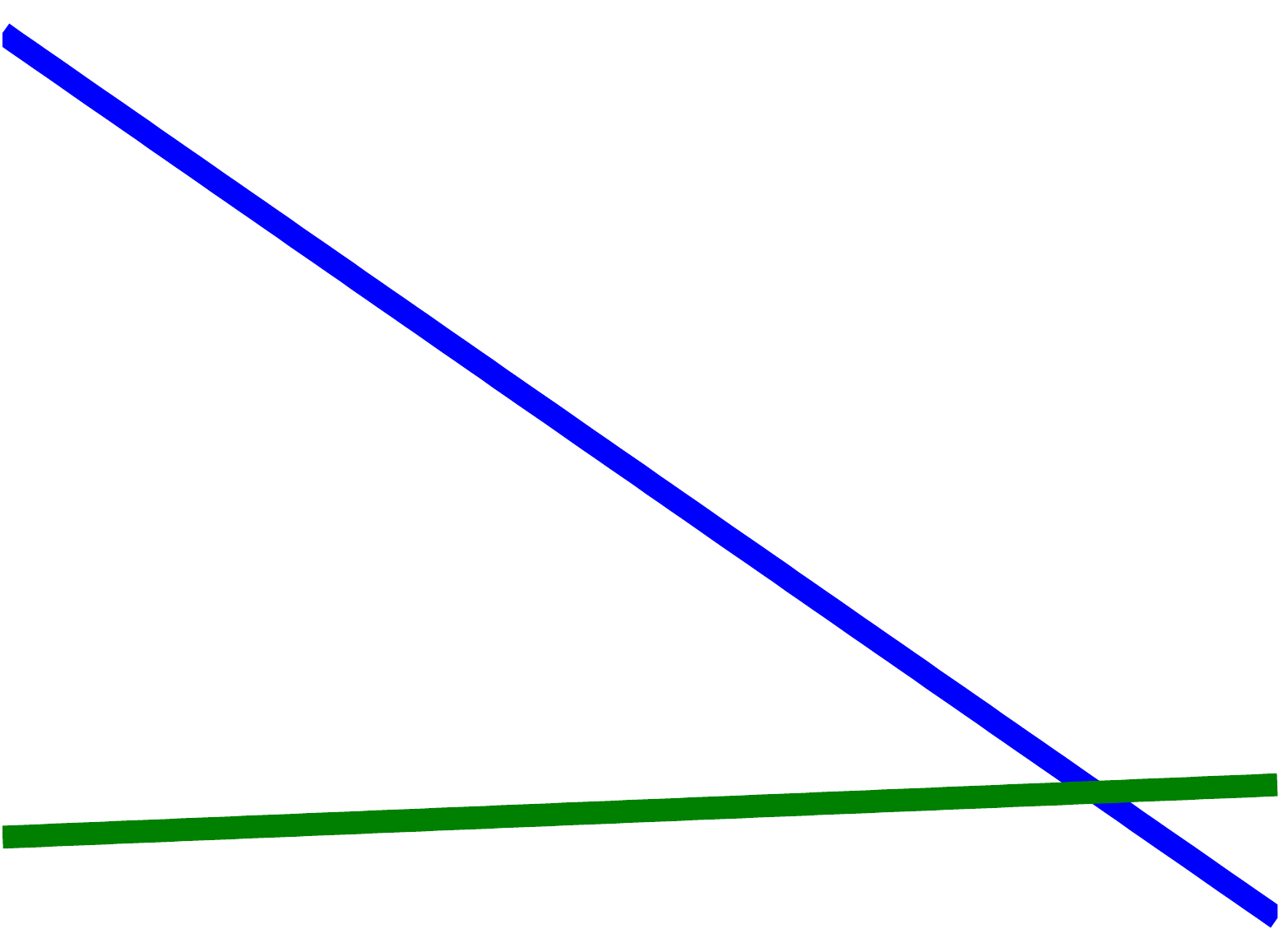} & 
\includegraphics[width=1.8cm,height=1.3cm]{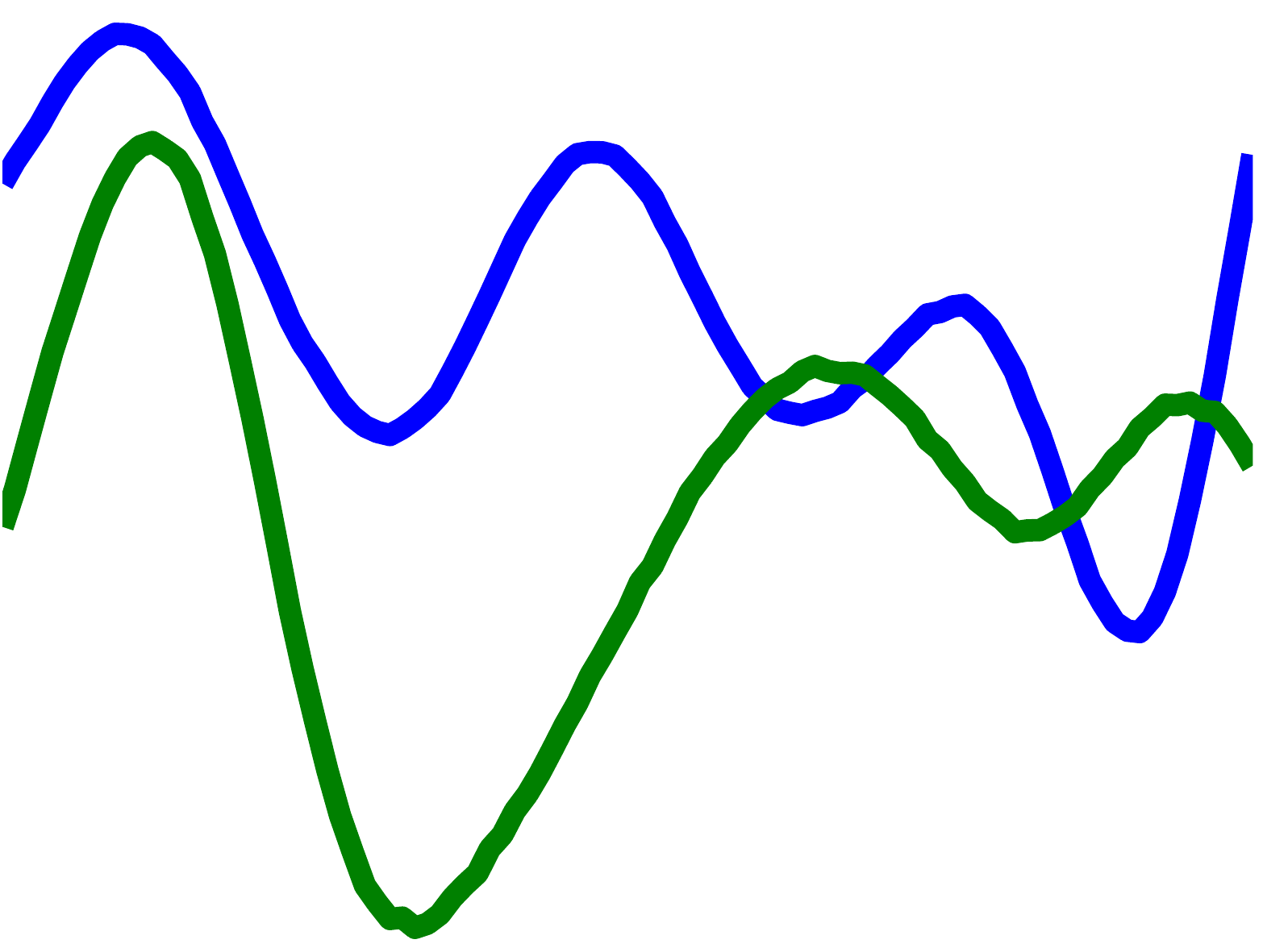} &
\includegraphics[width=1.8cm,height=1.3cm]{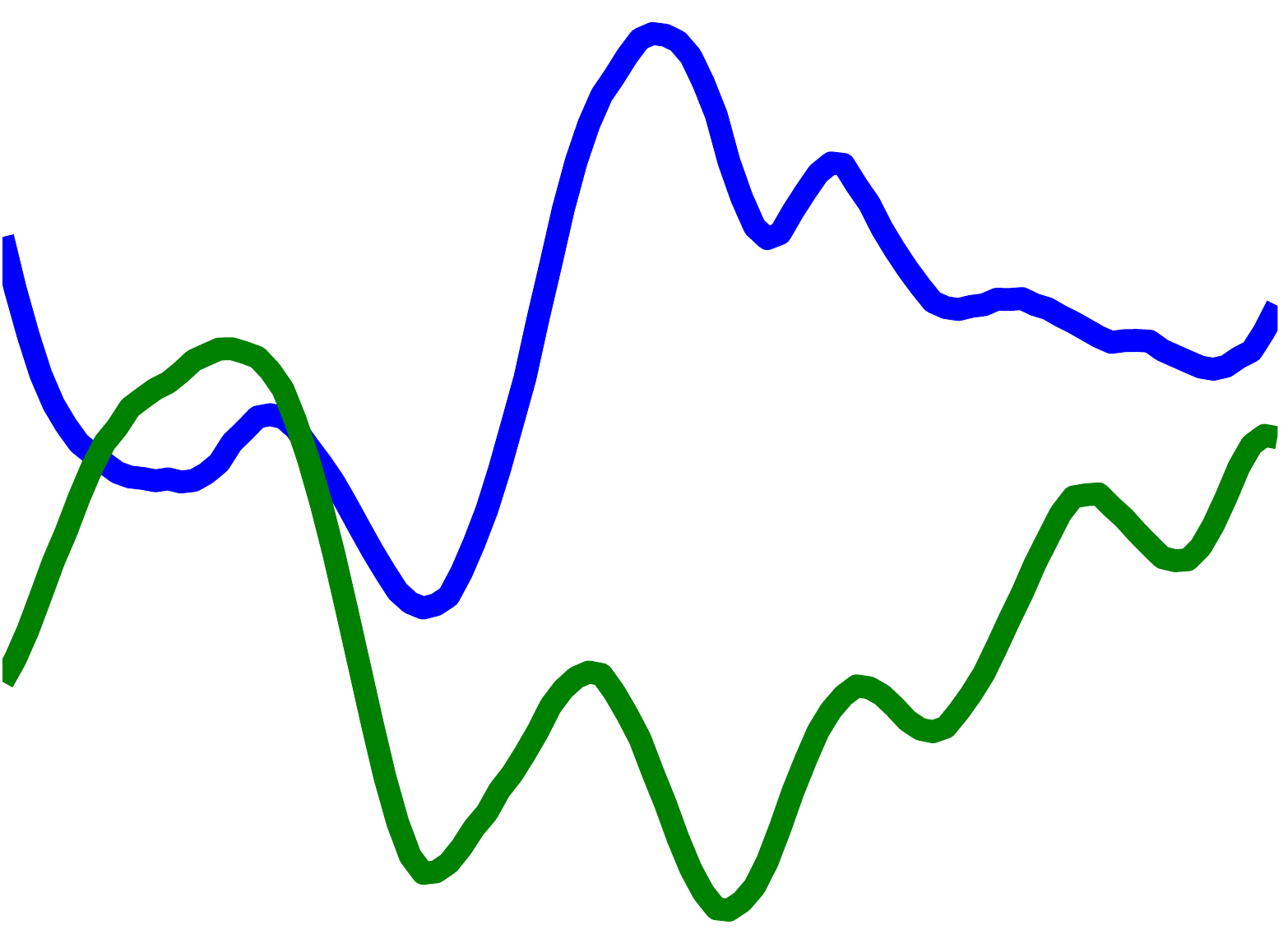} & 
\includegraphics[width=1.8cm,height=1.3cm]{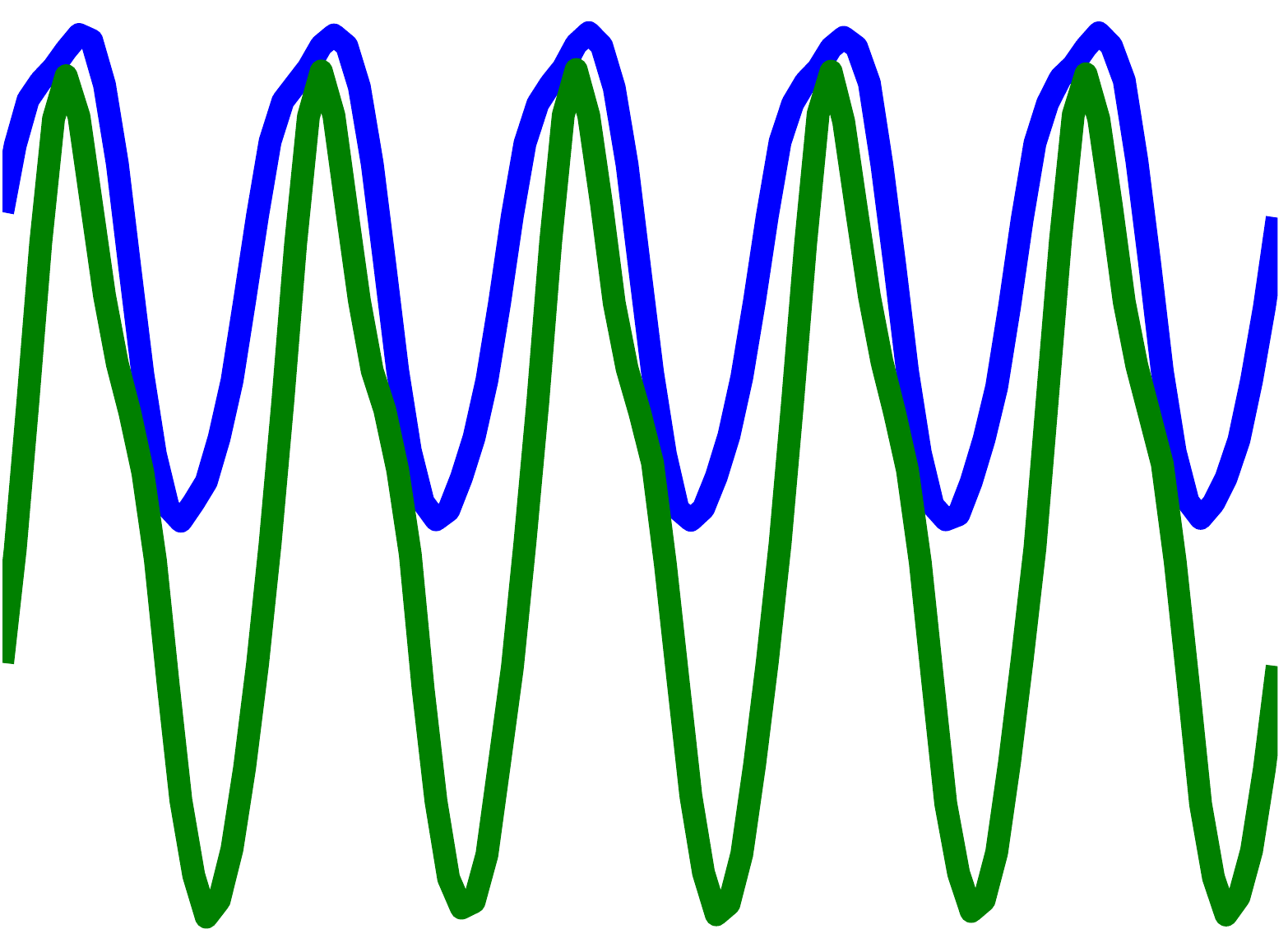} &
\includegraphics[width=1.8cm,height=1.3cm]{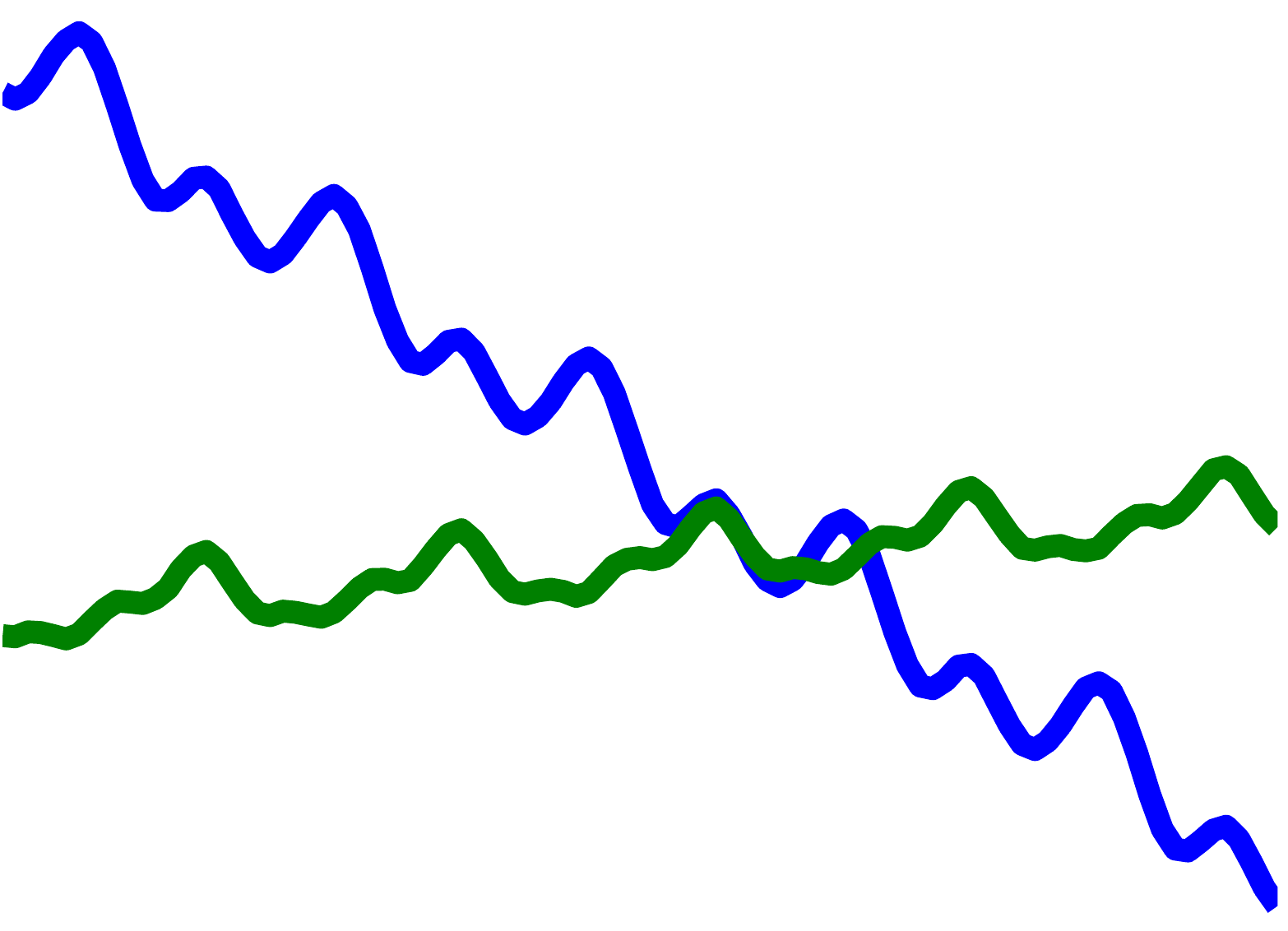} &
\includegraphics[width=1.8cm,height=1.3cm]{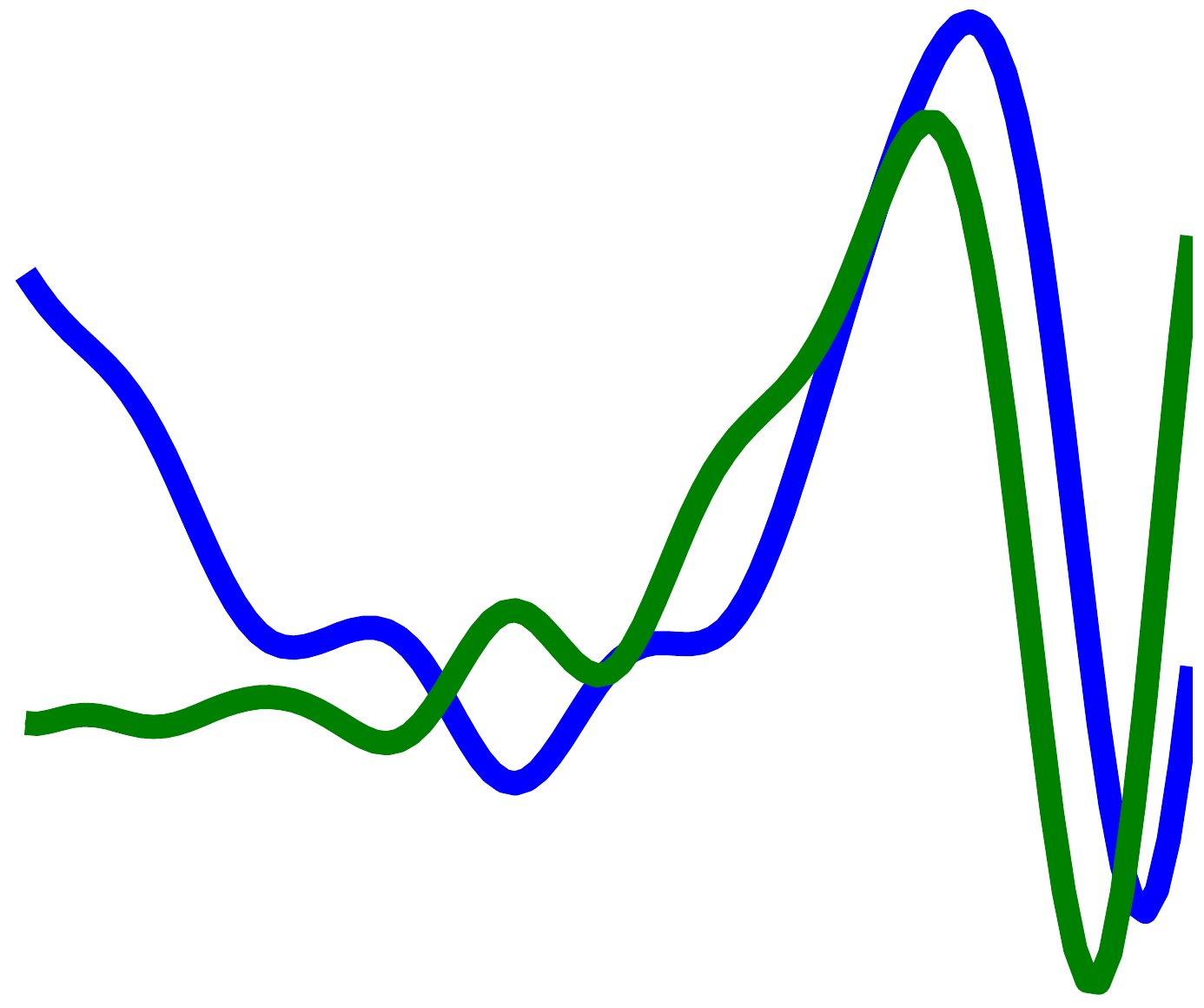}
\end{tabular}
\end{center}
\caption{{Base kernels (top) and two random draws from a GP with each respective kernel and combination of kernels (bottom).} \label{covariances1}}
\end{figure}

Commonly used kernels families include the squared exponential (SE), periodic (Per), linear (Lin), and rational quadratic (RQ), cf. Table~\ref{kernelitos}. Illustration of the base kernel and drawings from the GP prior is shown in Fig.~\ref{covariances1}. These base kernels can be actually combined following simple operations: summation, multiplication or convolution. {This way one may build sophisticated covariances from simpler ones. Note that the same essential property of kernel methods apply here: a valid covariance function must be positive semidefinite.} In general, the design of the kernel should rely on the information that we have for each estimation problem and should be designed to get the most accurate solution with the least amount of samples.

In Fig.~\ref{covariances1}, all the base kernels are one-dimensional. Nevertheless, kernels over multidimensional inputs can be actually constructed by adding and multiplying kernels over individual dimensions: (a) linear, (b) squared exponential (or RBF), (c) rational quadratic, and (d) periodic. See Table~\ref{kernelitos} for the explicit functional form of each kernel. Some simple kernel combinations are represented in the two last columns of the figure: a linear plus periodic covariances may capture strucutres that are periodic with trend (e), while a linear plus squared exponential covariances can accommodate structures with increasing variation (f).  By summing kernels, we can model the data as a superposition of independent functions, possibly representing different structures in the data. For example, in multitemporal image analysis, one could for instance dedicate a kernel for the time domain (perhaps trying to capture trends and seasonal effects) and another kernel function for the spatial domain (equivalently capturing spatial patterns and auto-correlations). In time series models, sums of kernels can express superposition of different processes, possibly operating at different scales: very often changes in geophysical variables through time occur at different temporal resolutions (hours, days, etc.), and this can be incorporated in the prior covariance with those simple operations. In multiple dimensions, summing kernels gives additive structure over different dimensions, similar to generalized additive models~\cite{Hastie09}. Alternatively, multiplying kernels allows us to account for interactions between different input dimensions or different notions of similarity. In the following section, we show how to design kernels that incorporate particular time resolutions, trends and periodicities.

\subsection{Time-based covariance for GPR}

Signals to be processed typically show particular characteristics, with time-dependent cycles and trends. One could include time $t_i$ as an additional feature in the definition of the input samples.  This {\em stacked approach}~\cite{
campsvalls06grsl} essentially relies on a covariance function $k({\bf z}_i,{\bf z}_j)$, where ${\bf z}_i=[t_i,\x_i]^\top$. The shortcoming is that the time relations are naively left to the nonlinear regression algorithm, and hence no explicit time structure model is assumed. To cope with this, one can use a linear combination (or composite) of different kernels: one dedicated to capture the different temporal characteristics, and the other to the feature-based relations. 

The issue here is how to design kernels capable of dealing with non-stationary processes. A possible approach is to use a {\em stationary} covariance operating on the variable of interest after being mapped with a nonlinear function engineered to discount such undesired variations. This approach was used in~\cite{Sampson92} to model {\em spatial patterns} of solar radiation with GPR. It is also possible to adopt a squared exponential (SE) as stationary covariance acting on the {\em time} variable mapped to a two-dimensional {\em periodic space} ${\bf z}(t)=[\cos(t),\sin(t))]^\top$, as explained in~\cite{Rasmussen06},
\begin{equation}
k(t_i,t_j) = \exp\bigg(-\dfrac{\|{\bf z}(t_i)-{\bf z}(t_j)\|^2}{2\sigma_t^2}\bigg),
\end{equation}
which gives rise to the following periodic covariance function
\begin{equation}
k(t_i,t_j) = \exp\bigg(-\dfrac{2\sin^2[(t_i-t_j)/2]}{\sigma_t^2}\bigg),
\end{equation}
where $\sigma_t$ is a hyper-parameter characterizing the periodic scale and needs to be inferred.
It is not clear, though, that the seasonal trend is exactly periodic, so we modify this equation by taking the product with a squared exponential component, to allow a decay away from exact periodicity:
\begin{equation}
k_2(t_i,t_j) = \gamma \exp\bigg(-\dfrac{2\sin^2[\pi(t_i-t_j)]}{\sigma_t^2}-\dfrac{(t_i-t_j)^2}{2\sigma_d^2}\bigg),
\end{equation}
where $\gamma$ gives the magnitude, $\sigma_t$ the smoothness of the periodic component, $\sigma_d$ represents the {\em decay-time} for the periodic component, and the period has been fixed to one year. Therefore, our final covariance is expressed as
\begin{equation}
k([\x_i, t_i],[\x_j, t_j]) = k_1(\x_i,\x_j) + k_2(t_i,t_j),
\end{equation}
which is parameterized by only three more hyperparameters collected in $\vect{\theta}=\{\nu,\sigma_1,\ldots,$ $\sigma_F,\sigma_n,\sigma_t,\sigma_d,\gamma\}$. Note that this kernel function allows us to incorporate time easily, but the relations between time $t_i$ and signal $\x_i$ samples is missing. Some approximations to deal with this issue exist in the literature, such as cross-kernel composition~\cite{campsvalls06grsl,campsvalls07composrvm} or latent force models~\cite{AlvarezLL13}.

We show the advantage of encoding such prior knowledge and structure in the relevant problem of solar irradiation prediction using GPR. Noting the non-stationary temporal behaviour of the signal, we develop a particular time-based composite covariance to account for the relevant seasonal signal variations. Data from the AEMET radiometric observatory of Murcia (Southern Spain, 38.0$^\circ$ N, 1.2$^\circ$ W) were used. Table~\ref{results} reports the obtained results with GPR models and several statistical regression methods: regularized linear regression (RLR), support vector regression (SVR), relevance vector machine (RVM) and GPR. All methods were run with and without using two additional dummy time features containing the year and day-of-year (DOY). We will indicate the former case with a subscript, like e.g. SVR$_t$.  First, including time information improves all baseline models. Second, the best overall results are obtained by the GPR models, when including time information or not. Third, in particular, the proposed temporal GPR (TGPR) outperforms the rest in accuracy (root-mean-square error, RMSE, and mean absolute error, MAE) and goodness-of-fit ($R$), and closely follows the elastic net in bias (ME). TGPR performs better than GPR and GPR$_t$ in all quality measures.

\begin{table}[t!]

\begin{center}
\caption{Results for the estimation of the daily solar {irradiation} of linear and nonlinear regression models. Subscript METHOD$_t$ indicates that the METHOD includes time as input variable. Best results are highlighted in bold, the second best in italics. \label{results}}
\renewcommand{\tabcolsep}{6pt}
\begin{tabular}{|l|c|c|c|c|}
\hline
\bf METHOD & \bf ME  & \bf RMSE   & \bf MAE  & \bf R \\
\hline
\hline
RLR                & 0.27 & 4.42 & 3.51 & 0.76 \\
RLR$_t$            & 0.25 & 4.33 & 3.42 & 0.78 \\
\hline
SVR~\cite{Smola2004}                 & 0.54 & 4.40 & 3.35 & 0.77 \\
SVR$_t$            & 0.42 & 4.23 & 3.12 & 0.79 \\
\hline
RVM~\cite{Tipping00}                 & 0.19 & 4.06 & 3.25 & 0.80 \\
RVM$_t$            & 0.14 & 3.71 & 3.11 & 0.81 \\
\hline
GPR~\cite{Rasmussen06}               & 0.14 & 3.22 & 2.47 & {\em 0.88} \\
GPR$_t$            & {\em 0.13} & {\em 3.15} & {\em 2.27} & {\em 0.88} \\
{\bf TGPR}         & {\bf 0.11} & {\bf 3.14} & {\bf 2.19} & {\bf 0.90} \\
\hline
\end{tabular}
\end{center}
\end{table}

\subsection{Heteroscedastic GPR: Learning the noise model}

The standard GPR is essentially homoscedastic, i.e., assumes constant noise power $\sigma^2$ for all observations. This assumption can be too restrictive for some problems. Heteroscedastic GPs, on the other hand, let noise power vary smoothly throughout input space, by changing the prior over $\varepsilon_n$ to $\varepsilon_n \sim \Normal(0, e^{g(\x_n)})$, 
and placing a GP prior over  $g(\x)\;\sim\;\GP(\mu_0\vect{1}, k_{\vect{\theta}_g}(\vect{x},\vect{x}'))$. Note that the exponential is needed\footnote{{Of course, other transformations are possible, just not as convenient.}} in order to describe the non-negative variance. The hyperparameters of the covariance functions of both GPs are collected in $\vect{\theta}_f$ and $\vect{\theta}_g$, accounting for the signal and the noise relations, respectively.

Relaxing the homoscedasticity assumption into heteroscedasticity yields a richer, more flexible model that contains the standard GP as a particular case corresponding to a constant $g(\x)$. Unfortunately, this also hampers analytical tractability, so approximate methods must be used to obtain posterior distributions for $f(\x)$ and $g(\x)$, which are in turn required to compute the predictive distribution over $y_*$. 

\begin{figure}[h!]
\centerline{
\includegraphics[width=7.5cm]{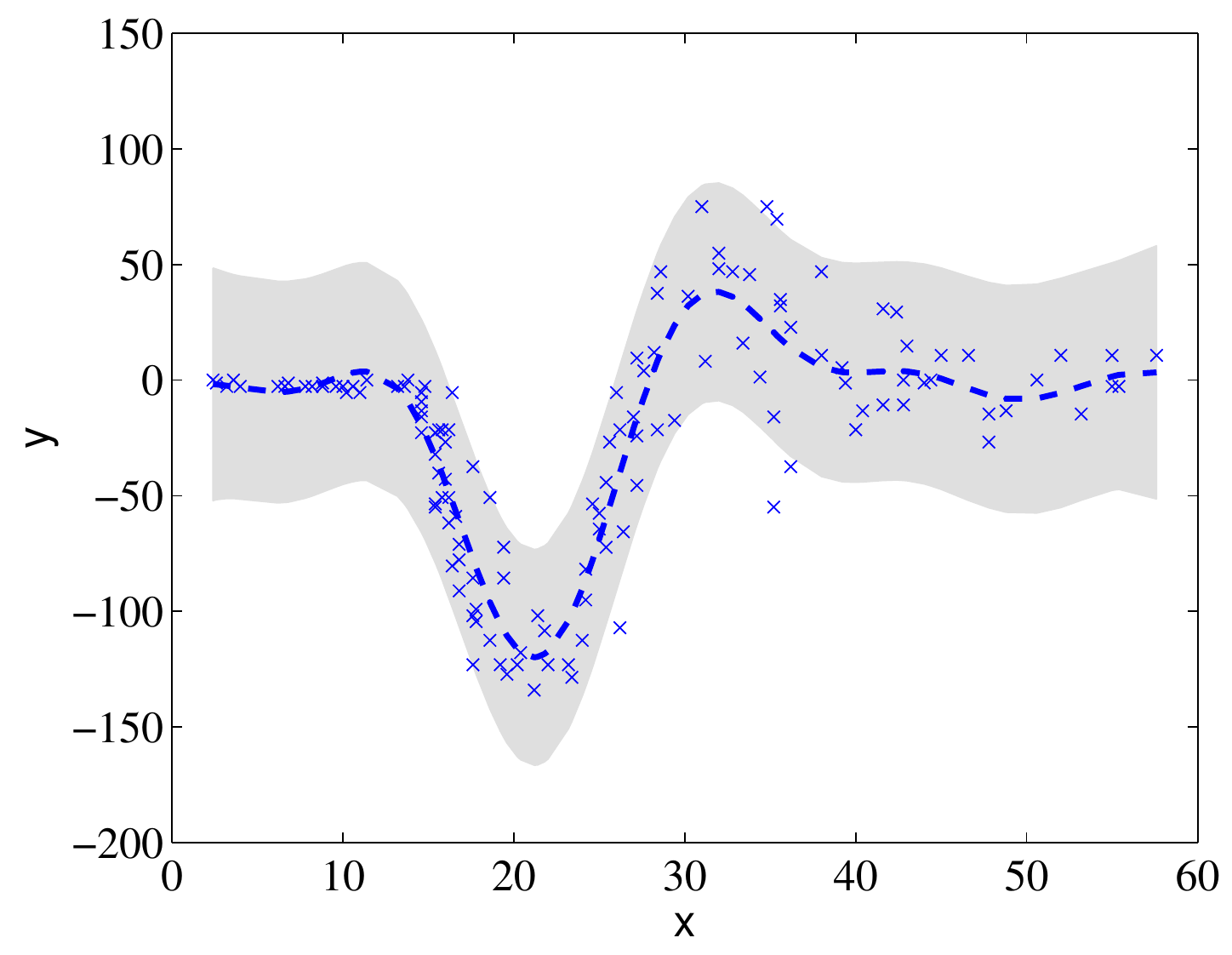}
\includegraphics[width=7.5cm]{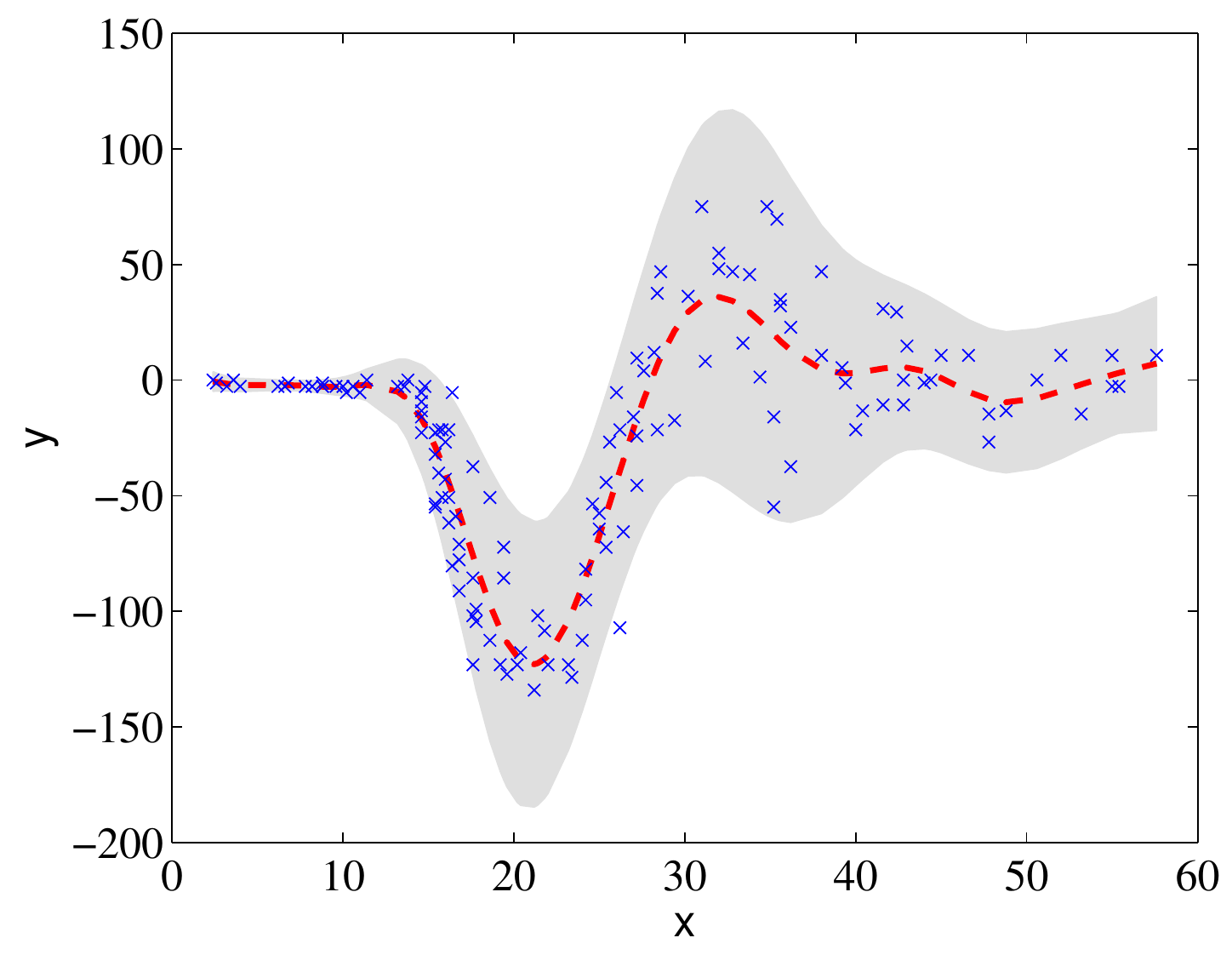}
}
\caption{Predictive mean and variance of the standard GP (left) and the heteroscedastic GP (right). It is noticeable that in the low noise regime the VHGP produces tighter confidence intervals as expected, while high noise variance associated to high signal variance (middle of the observed signal) the predictive variance is more reasonable too. \label{homohetero}}
\end{figure}

As an alternative to the costly classic heteroscedastic GP approaches, variational techniques allow to approximate intractable integrals arising in Bayesian inference and machine learning. A sophisticated variational approximation called \emph{Marginalized Variational (MV)} approximation was introduced in \cite{Lazaro11}. The MV approximation renders (approximate) Bayesian inference in the heteroscedastic GP model both fast and accurate. We will refer to this variational approximation for heteroscedastic GP regression as VHGPR. A simple comparison between the homoscedastic canonical GP and the VHGPR model is shown in Fig.~\ref{homohetero}.

\subsection{Warped GPR: Learning the output transformation}

Very often, in practical applications, one transforms the observed variable to better pose the problem. Actually, it is a standard practice to `linearize' or `uniformize' the distribution of the observations (which is commonly skewed due to the sampling strategies in {\em in situ} data collection) by applying non-linear link functions like the logarithmic, the exponential or the logistic functions. 

{\em Warped} GPR~\cite{Snelson03} essentially warps observations ${\bf y}$ through a nonlinear parametric function $g$ to a latent space $z_i= g(y_i) = g(f({\bf x}_i) + \varepsilon_i)$, where $f$ is a possibly noisy latent function with $d$ inputs, and $g$ is a function with scalar inputs parametrized by $\vect{\psi}$. The function $g$ must be {\em monotonic}, otherwise the probability measure will not be conserved in the transformation, and the distribution over the targets may not be valid~\cite{Snelson03}. It can be shown that replacing $y_i$ by $z_i$ into the standard GP model leads to an extended problem that can be solved by taking derivatives of the negative log likelihood function in~\eqref{eq:logevidence}, but now with respect to both $\vect{\theta}$ and $\vect{\psi}$ parameter vectors.

For both the GPR and WGPR models we need to define the covariance (kernel, or Gram) function $k(\cdot,)$, which should capture the similarity between samples. We used the standard Automatic Relevance Determination (ARD) covariance~\cite{Rasmussen06}. Model hyperparameters are collectively grouped in $\vect{\theta}=\{\nu,\sigma_n,\sigma_1,\ldots,\sigma_d\}$. In addition, for the WGPR we need to define a parametric smooth and monotonic form for $g$, which can be defined as:
$$g(y_i;\vect{\psi}) = \sum_{\ell=1}^L a_\ell~\text{tanh}(b_\ell~y_i + c_\ell),~~~~a_\ell,~b_\ell\geq 0, $$ 
where $\vect{\psi}=\{{\bf a},{\bf b},{\bf c}\}$. Recently, flexible non-parametric functions have replaced such parametric forms~\cite{Lazaro12warp}, thus placing another prior for $g(\x)\;\sim\;\GP(f,c(f,f'))$, whose model is learned via variational inference.

For illustration purposes, we focus on the estimation of chlorophyll-a concentrations from remote sensing upwelling radiance just above the ocean surface. We used the SeaBAM dataset~\cite{oreilly98,maritorena00}, which gathers 919 {\em in situ} pigment measurements around the United States and Europe. The dataset contains coincident {\em in situ} chlorophyll concentration and remote sensing reflectance measurements (Rrs($\lambda$), [sr$^{-1}$]) at some wavelengths (412, 443, 490, 510 and 555 nm) that are present in the SeaWiFS ocean colour satellite sensor. The chlorophyll concentration values range from 0.019 to 32.79 mg/m$^3$ (revealing a clear exponential distribution).

\begin{table}[t!]
\begin{center}

\caption{Results using both raw and empirically-transformed observation variables.}\label{res}
\renewcommand{\tabcolsep}{4pt}
\begin{tabular}{|l|c|c|c|c|}
\hline
	 & \bf ME	 & \bf RMSE	 & \bf MAE	 & \bf R \\ 
\hline
\hline
\bf Raw &&&&\\
\hline
GPR	 & 0.02	 & 1.74	 & 0.33	 & 0.82 \\ 
VHGPR	 & 0.29	 & 2.51	 & 0.46	 & 0.65 \\ 
WGPR	 & 0.08	 & 1.71	 & 0.30	 & 0.83 \\ 
\hline
\hline
\bf Empirically-based &&&&\\
\hline
\hline
GPR	 & 0.15	 & 1.69	 & 0.29	 & 0.86 \\ 
VHGPR	 & 0.15	 & 1.70	 & 0.29	 & 0.85 \\ 
WGPR	 & 0.17	 & 1.75	 & 0.30	 & 0.86 \\ 
\hline
\end{tabular}
\end{center}
\end{table} 

Table~\ref{res} shows different scores --bias (ME), accuracy (RMSE, MAE) and goodness-of-fit ($R$)-- between the observed and predicted variable when using the raw data (no {\em ad hoc} transform at all) and the empirically adjusted transform. Results are shown for three flavours of GPs: the standard GPR~\cite{Rasmussen06}, the variational heteroscedastic GP (VHGPR)~\cite{CampsVallsGRSL2013}, and the proposed warped GP regression (WGPR)~\cite{Snelson03,Lazaro12warp} for different rates of training samples. Empirically-based warping slightly improves the results over working with raw data for the same number of training samples, but this requires prior knowledge about the problem, time and efforts to fit an appropriate function. On the other hand, WGPR outperforms the rest of GPs in all comparisons over standard GPR and VHGPR ($\sim +1-10\%$). Finally, WGPR nicely compensates the lack of prior knowledge about the (possibly skewed) distribution of the observation variable.

\subsection{Source code and toolboxes}

The most widely known sites to obtain free source code on GP modeling are GPML\footnote{\url{http://www.gaussianprocess.org/}} and GPstuff\footnote{\url{http://becs.aalto.fi/en/research/bayes/gpstuff/}}. The former website centralizes the main activities in GP modeling and provides up-to-date resources concerned with probabilistic modeling, inference and learning based on GPs, while the latter is a versatile collection of GP models and computational tools required for inference, sparse approximations and model assessment methods. We also recommend to the interested reader in regression in general, our MATLAB SimpleR\footnote{\url{http://isp.uv.es/soft.htm}} toolbox that contains many regression tools organized in families: tree-based, bagging and boosting, neural nets, kernel regression methods, and several Bayesian nonparametric models like GPs.

\section{Analysis of Gaussian Process Models}\label{sec:analysis}

An interesting possibility in GP models is to extract knowledge from the trained model. We will show in what follows two different approaches: 1) feature ranking exploiting the automatic relevance determination (ARD) covariance and 2) uncertainty estimation looking at the predictive variance estimates.

\subsection{Ranking features through the ARD covariance}

\begin{figure}[h!]
\centerline{
\IG[height=5cm]{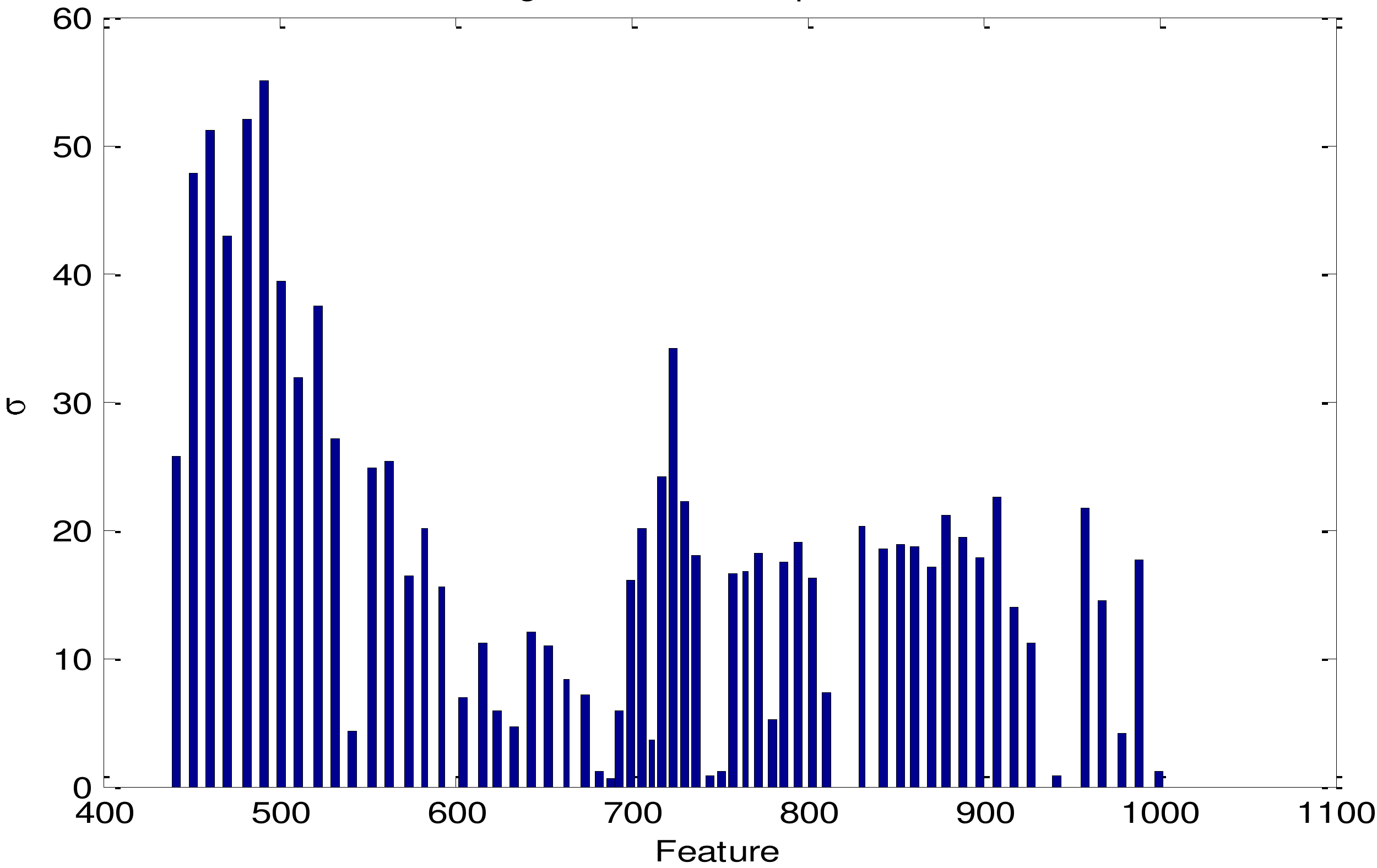}~~~
\IG[height=5cm]{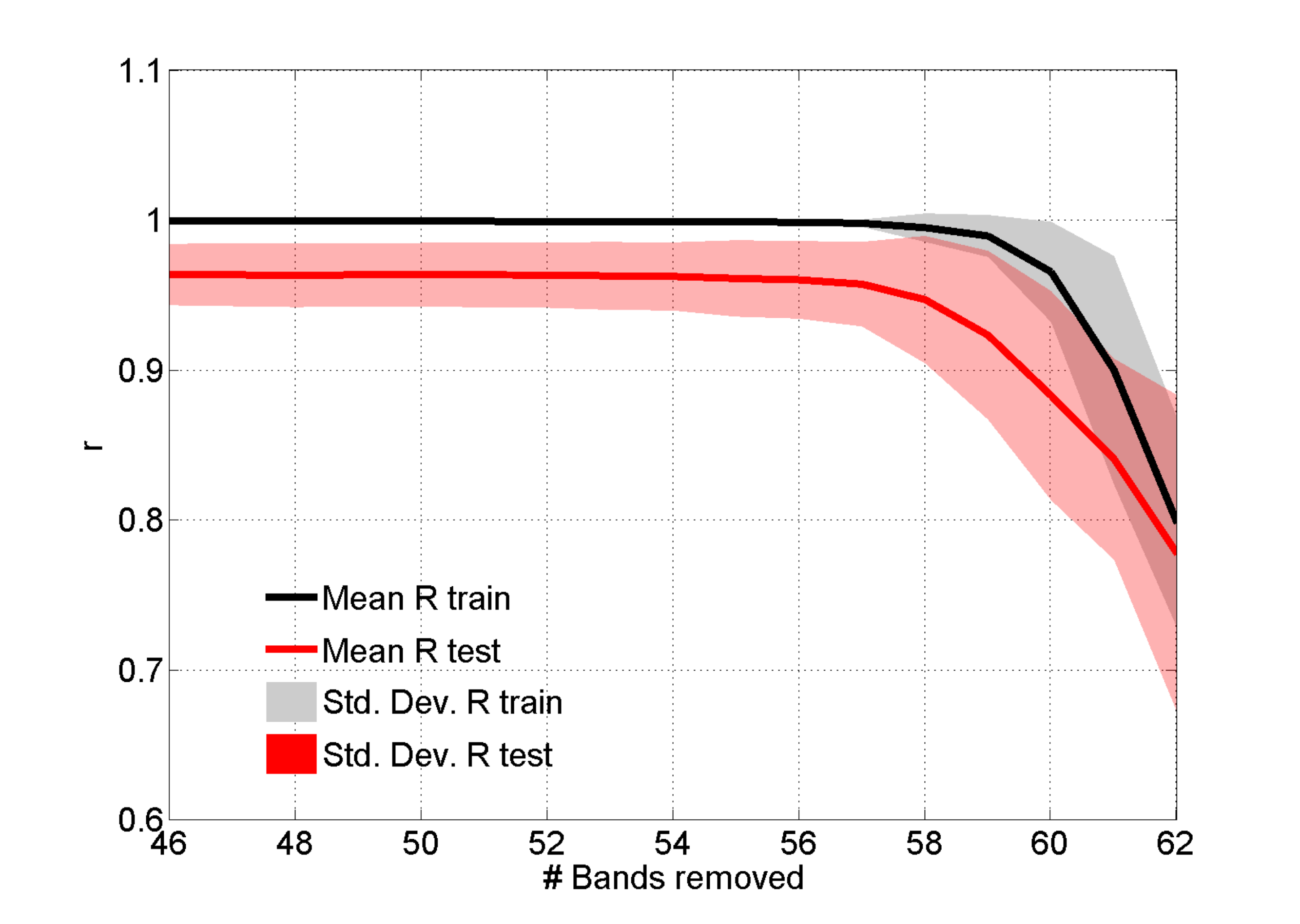}\\
}
\caption{Estimated $\sigma_b$ values for one GP model using 62 CHRIS bands (left). The lower the $\sigma_b$ the more important the band is for regression. $Chl$ $r$ and standard deviation (SD) of training and validation for GP fittings using backward elimination (right).}
\label{Fsigma}
\end{figure}

One of the advantages of GPs is that during the development of the GP model the predictive power of each single band is evaluated for the parameter of interest through calculation of the ARD. Specifically, band ranking through $\sigma_b$ may reveal the bands that contribute the most to the development of a GP model. An example of the $\sigma_b$'s for one GP model trained with field leaf chlorophyll content ($Chl$) data and with 62 CHRIS bands is shown in Fig.~\ref{Fsigma} (left). The band with highest $\sigma_b$ is the least contributing to the model. It can be noted that a relatively few bands (about 8) were evaluated as crucial for $Chl$ estimation, while the majority of bands were evaluated as less contributing.

This does not necessarily mean that other bands are obstructing optimized accuracies. Only when less than 4 bands were left accuracies started to degrade rapidly Fig.~\ref{Fsigma} (right). The figure suggests that the most relevant spectral region is to be found between 550 and 1000 nm. Most contributing bands were positioned around the red edge, at 680 and 730 nm respectively, but not all bands within the red edge were evaluated as relevant. This is due to when having a large number of bands available then neighbouring bands do not provide much additional information and can thus be considered as redundant. 

\begin{figure}[h!]
\begin{center}
\IG[width=10.5cm]{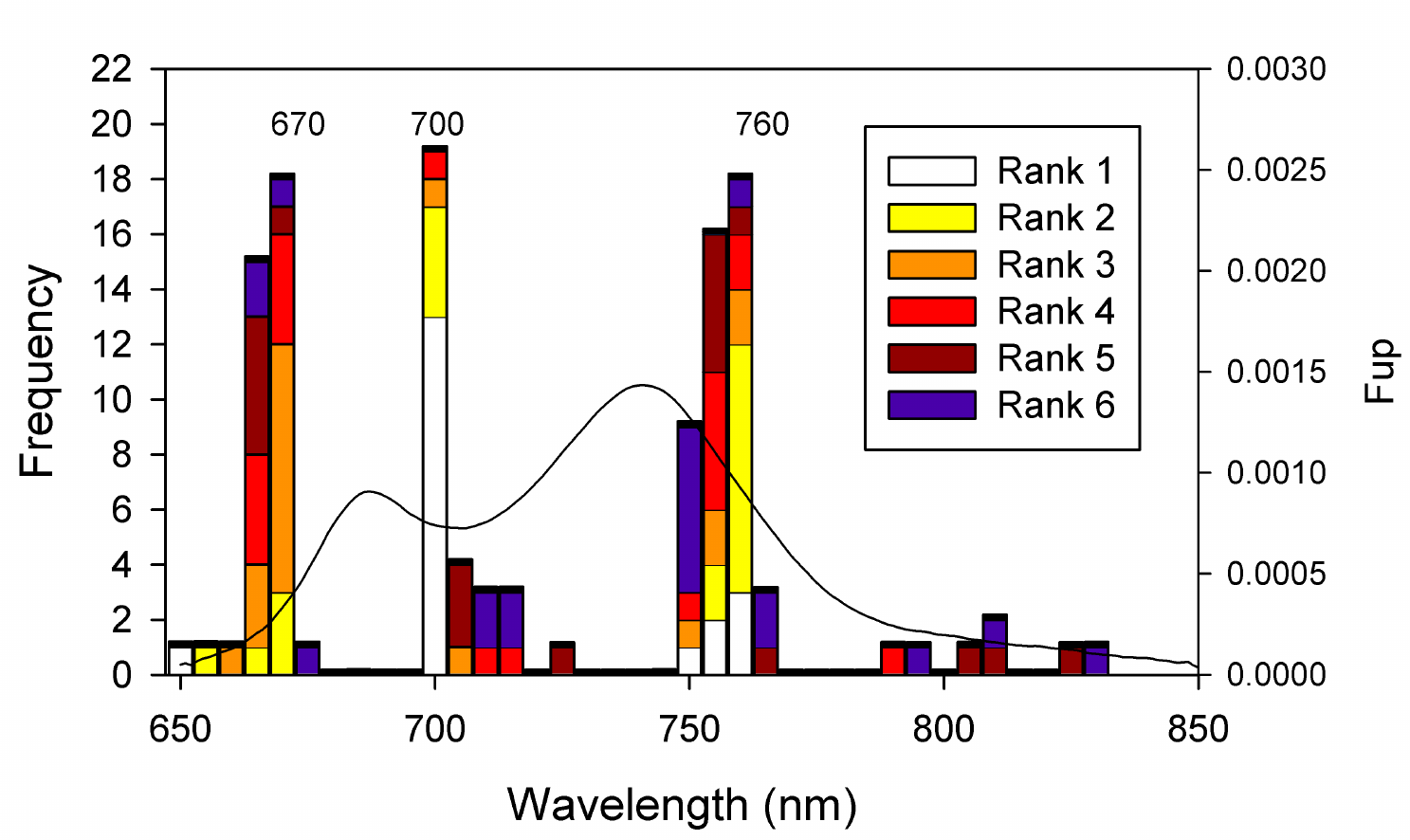}
\end{center}
\vspace{-0.7cm}
\caption{Frequency plots of the top eight ranked bands with lowest $\sigma_b$  values in 20 runs of GPR prediction of $Chl$  based on upward fluorescence ($F_{up}$) emission. An emission curve is given as illustration.}
\label{Fsigma_Shari}
\end{figure}

Consequently, the $\sigma_b$ proved to be a valuable tool to detect most sensitive bands of a sensor towards a biophysical parameter. A more systematic analysis was applied by sorting the bands on their relevance and counting the band rankings over 50 repetitions. In~\cite{Verrelst12rse} the four most relevant bands were tracked for $Chl$, LAI and fCOVER and for different Sentinel-2 settings. It demonstrated the potential of Sentinel-2, with its new band in the red-edge, for vegetation properties estimation. Also in~\cite{Verrelst2015} $\sigma_b$ were used to analyze band sensitivity of Sentinel-2 towards LAI. A similar approach was pursued on analyzing leaf $Chl$ based on tracking the most sensitive spectral regions of sun-induced fluorescence data~\cite{VanWittenberghe2014}, as displayed in Fig.~\ref{Fsigma_Shari}.

\subsection{Uncertainty intervals}

In this section, we use GP models for retrieval and portability in space and time. For this, we will exploit the associated predictive variance (i.e. uncertainty interval) provided by GP models. Consequently, retrievals with high uncertainties refer to pixel spectral information that deviates from what has been represented during the training phase. In turn, low uncertainties refer to pixels that were well represented in the training phase. The quantification of variable-associated uncertainties is a strong requirement when remote sensing products are ingested in higher level processing, e.g. to estimate ecosystem respiration, photosynthetic activity, or carbon sequestration~\cite{Jagermeyr2014}. 

\begin{figure}[t!]
\begin{center}

\setlength{\tabcolsep}{1mm}
\begin{tabular}{ccc}
{\em Chl} & {\em LAI} & {\em fCOVER} \\
\IG[width=3.8cm]{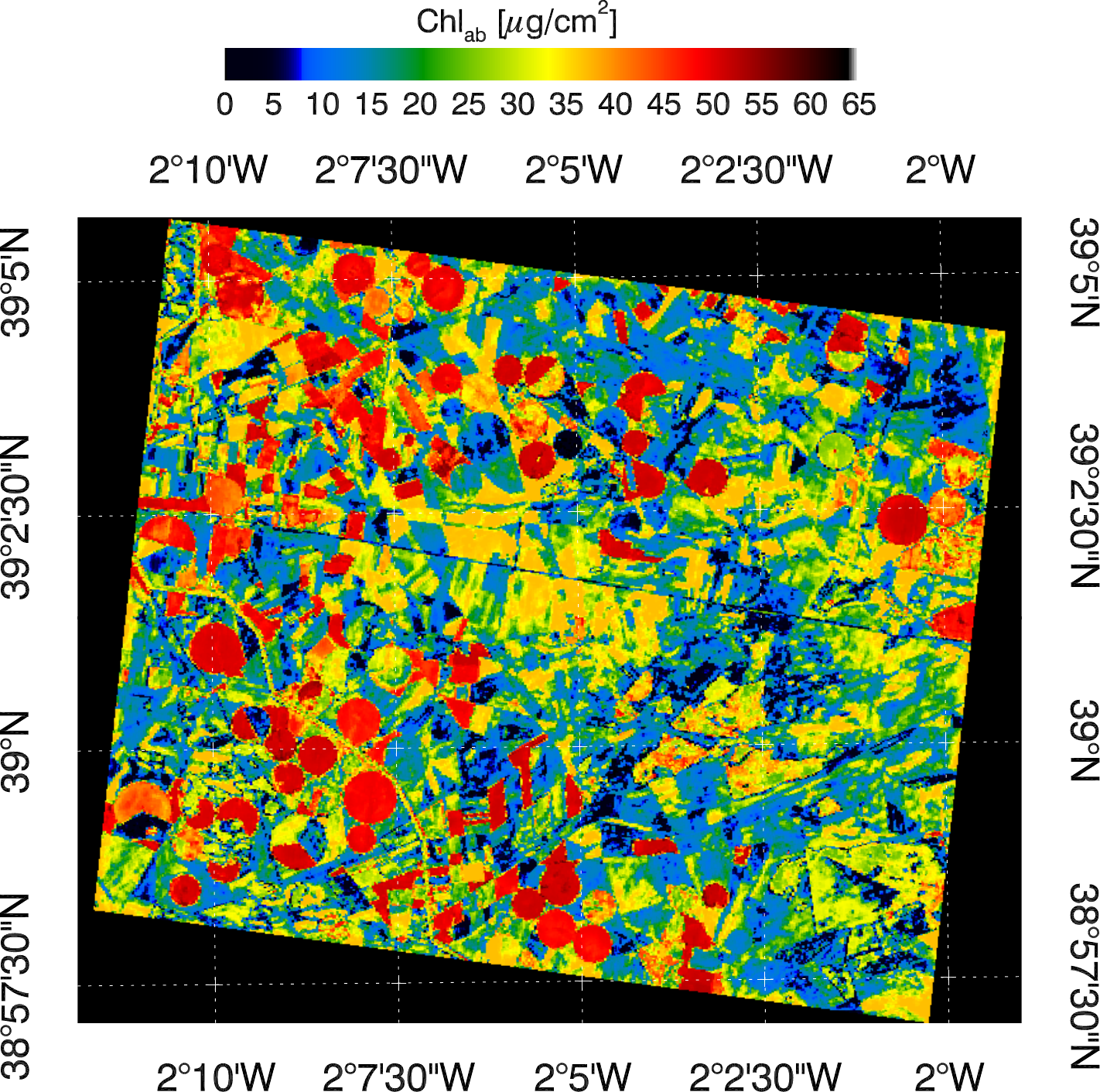} &  
\IG[width=3.8cm]{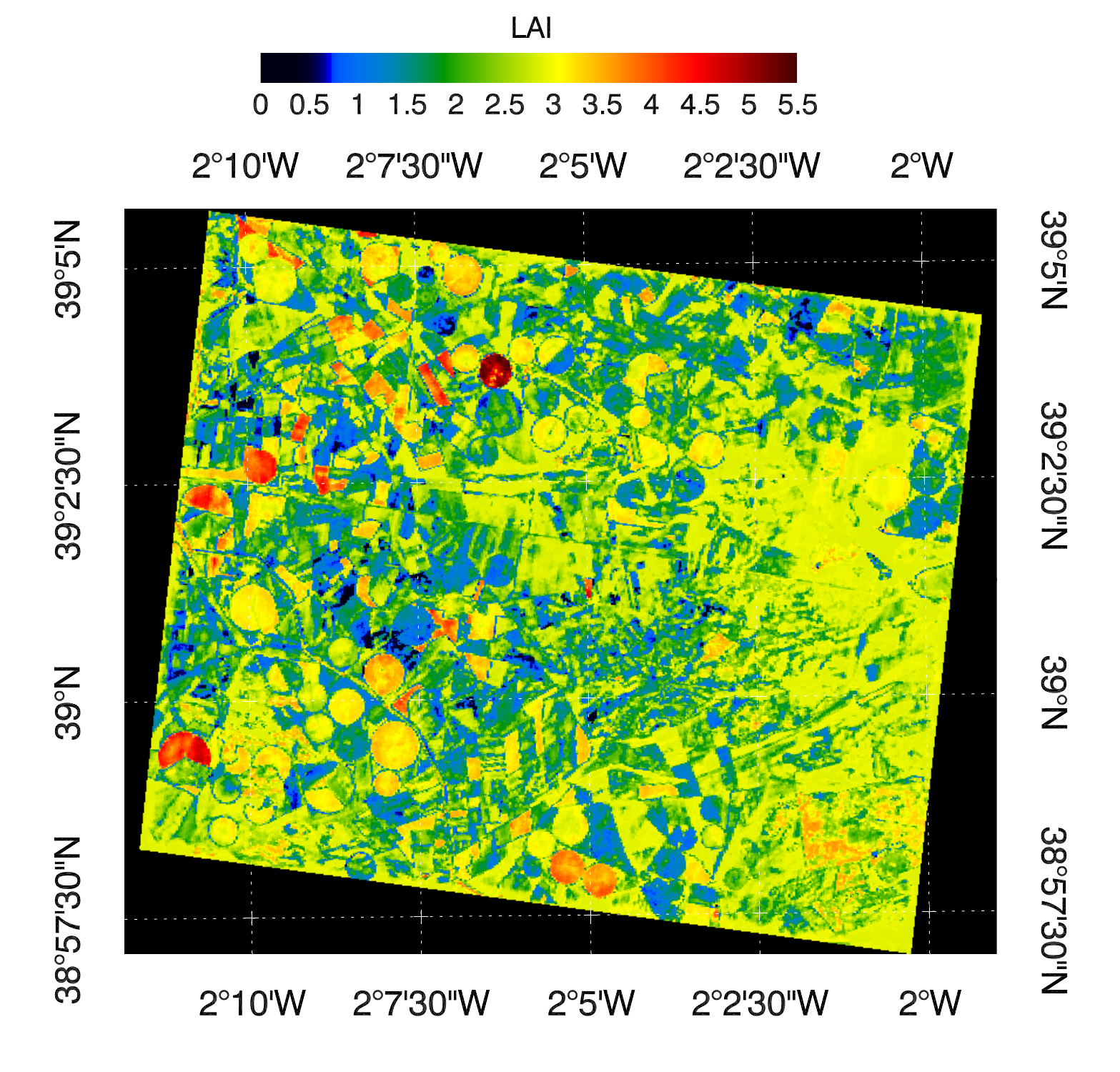} &  
\IG[width=3.8cm]{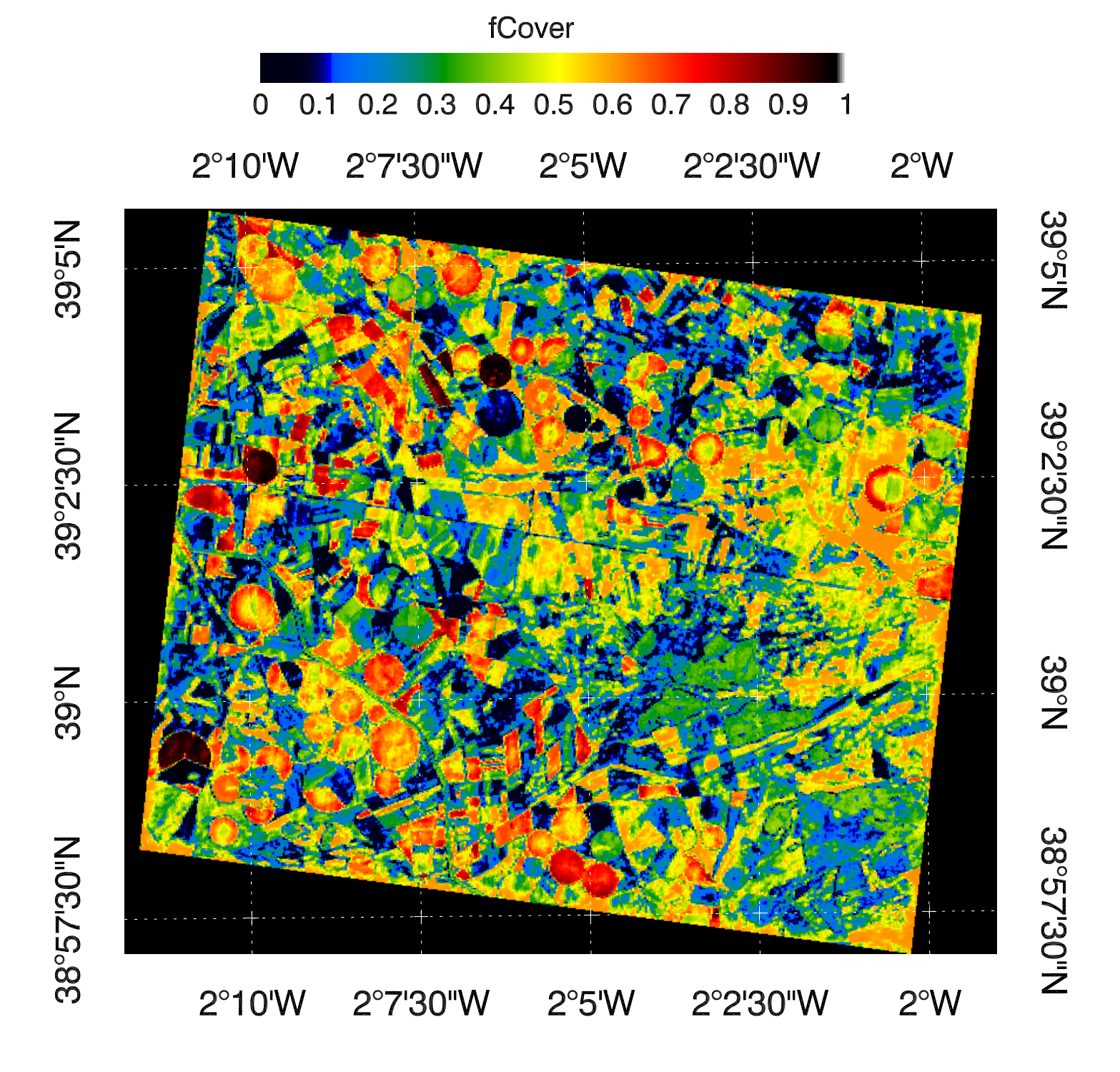} \\
\IG[width=3.8cm]{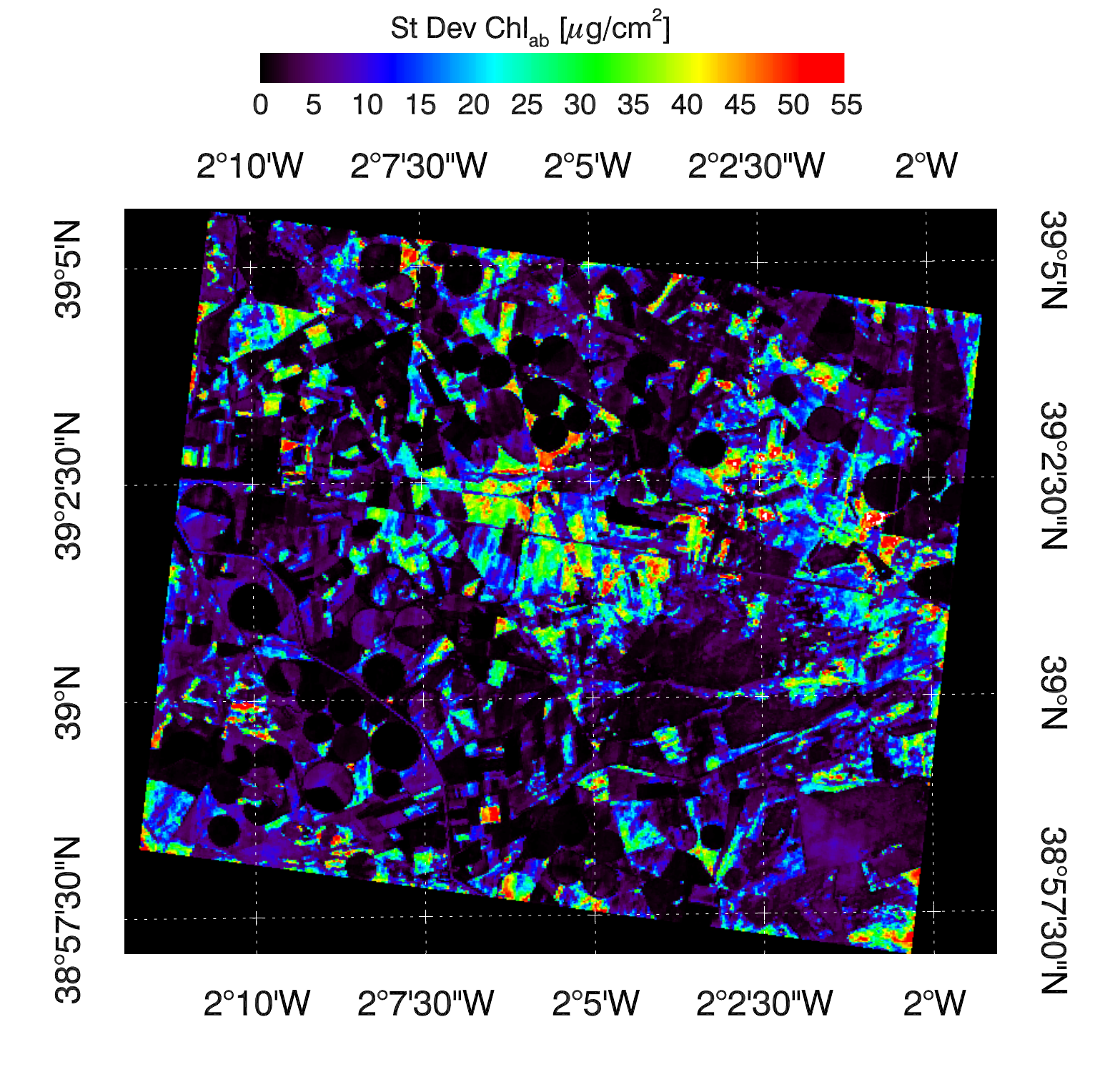} &  
\IG[width=3.8cm]{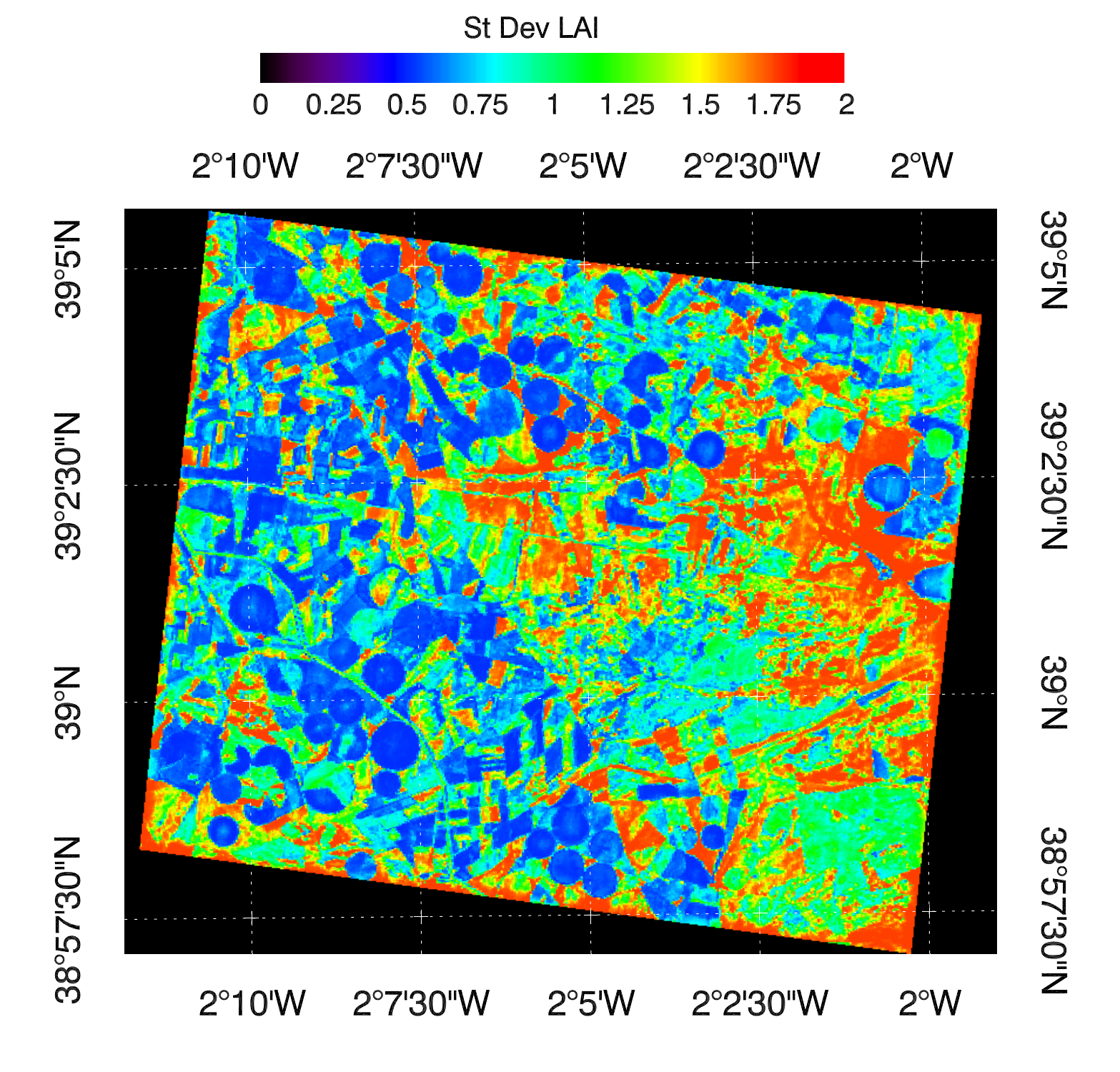} &  
\IG[width=3.8cm]{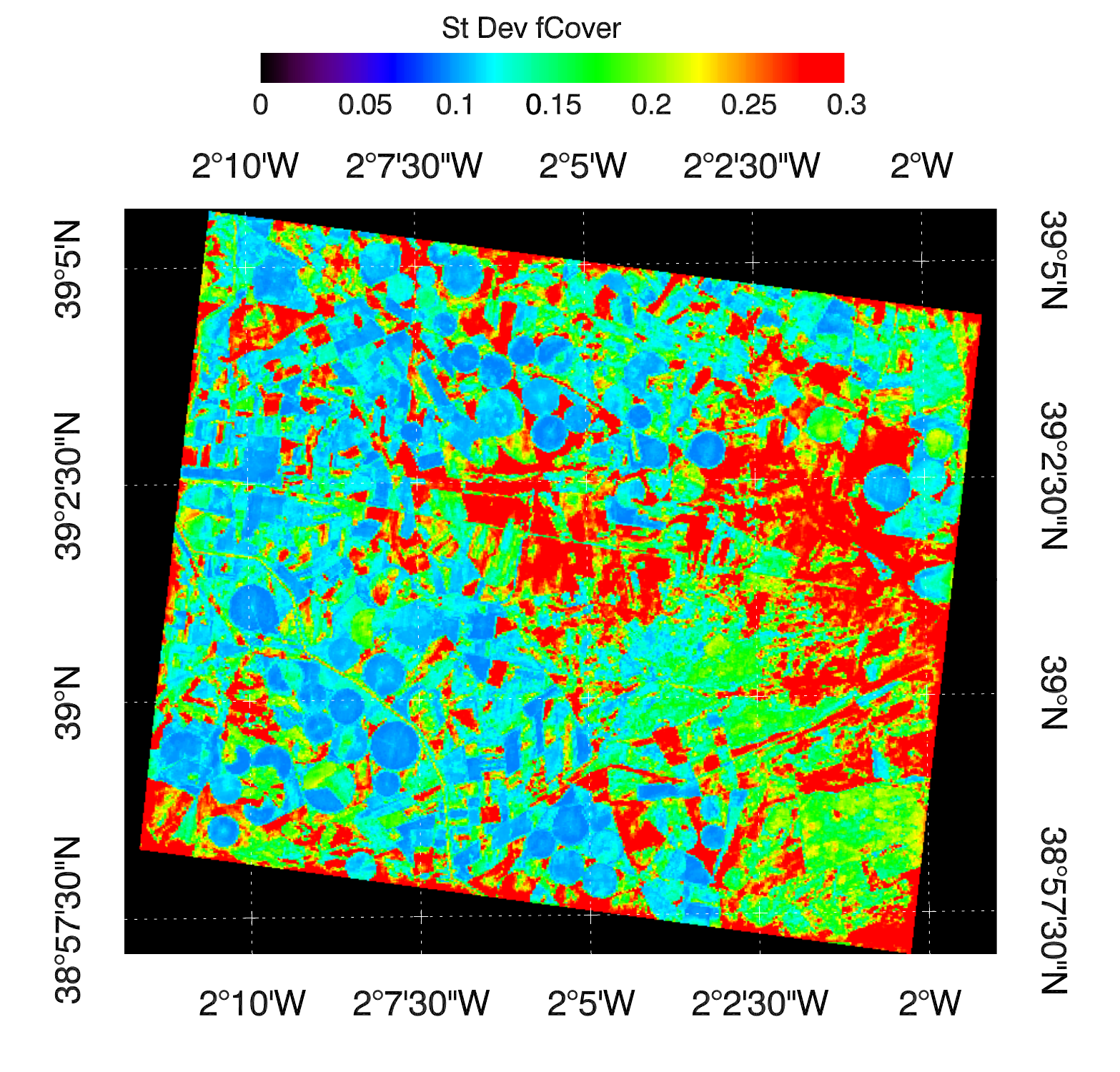} \\
\end{tabular}
\end{center}
\vspace{-0.5cm}
\caption{Prediction maps (top) and associated uncertainty intervals (bottom), generated with GP and four bands of the CHRIS 12-07-2003 nadir image.}
\label{GPRbiophysmaps}
\end{figure}

The application of GPs for the estimation of biophysical parameters was initially demonstrated in~\cite{Verrelst12b}. A locally collected field dataset called SPARC-2003  at Barrax (Spain) was used for training and validation of GPs for the vegetation parameters of LAI, $Chl$ and fCOVER. Sufficiently high validation accuracies were obtained ($R^{2}$ > 0.86) for processing a CHRIS image into these parameters, as shown in Fig.~\ref{GPRbiophysmaps}. Within the uncertainty maps, areas with reliable retrievals are clearly distinguished from areas with unreliable retrievals. Low uncertainties were found on irrigated areas and harvested fields. High uncertainties were found on areas with remarkably different spectra, such as bright, whitish calcareous soils, or harvested fields. This indicates that the input spectrum deviates from what has been presented during the training stage, thereby imposing uncertainties to the retrieval. 

GP models were subsequently applied to the SPARC dataset that was re-sampled to different Sentinel-2 band settings and then uncertainties were inspected~\cite{Verrelst12rse}. On the whole, adding spectral information led to reduction of uncertainties and thus more meaningful biophysical parameter maps. The locally-trained GP models were applied to simulated Sentinel-2 images in a follow-up study~\cite{Verrelst2013c}. Time series over the local Barrax site as well images across the world were processed. Also the role of an extended training dataset by adding spectra of non-vegetated surfaces were evaluated. Subsequently the uncertainty values were analyzed. By using the extended training dataset not only further improved performances but also allowed a decrease in theoretical uncertainties. The GP models were successfully applied to simulated Sentinel-2 images covering various sites; associated relative uncertainties were on the same order as those generated by the reference image. 

\begin{figure}[h!]
\begin{center}

\setlength{\tabcolsep}{-4mm}
\begin{tabular}{ccc} 
\IG[width=3.8cm]{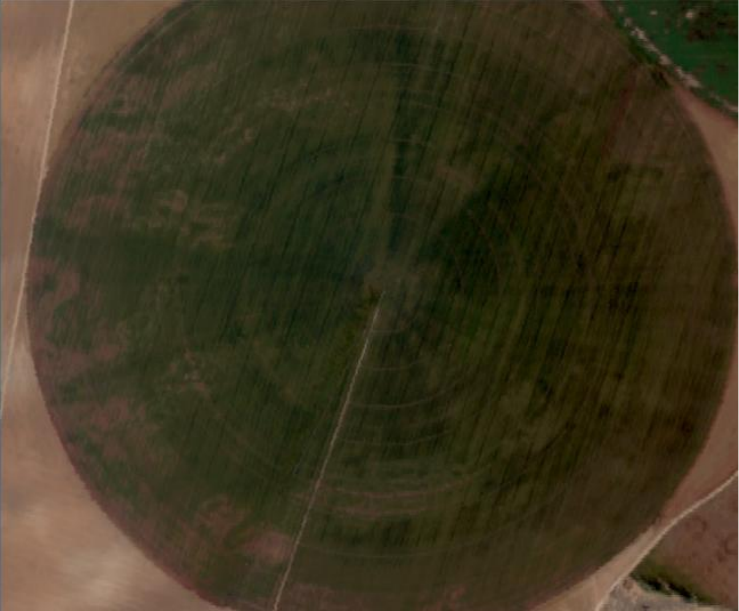} 
\IG[width=3.8cm]{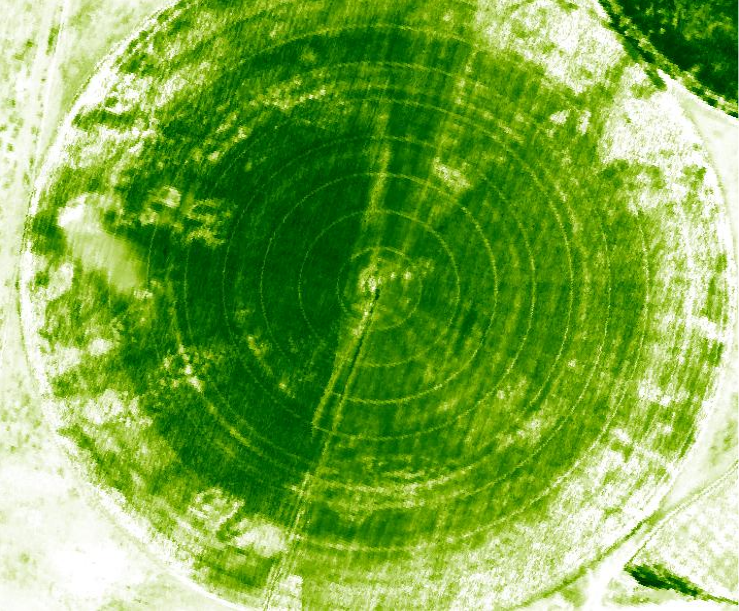} 
\IG[width=3.8cm]{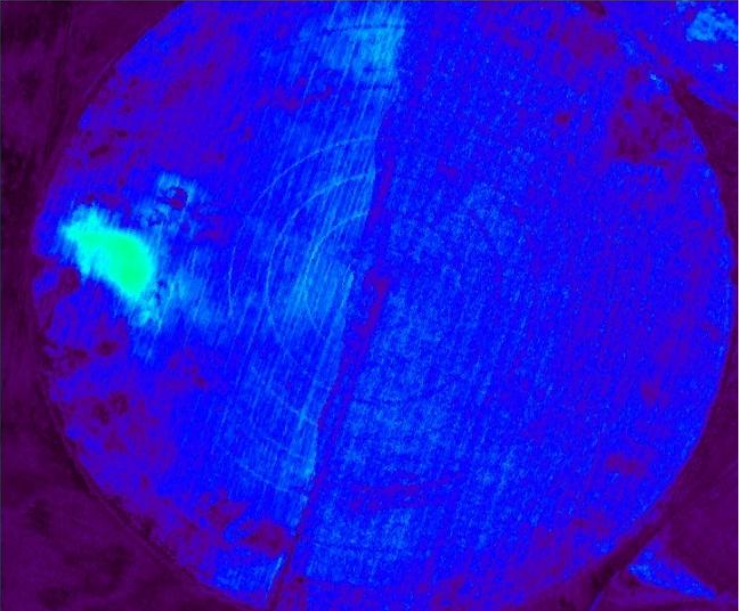}\\
\IG[width=3.8cm]{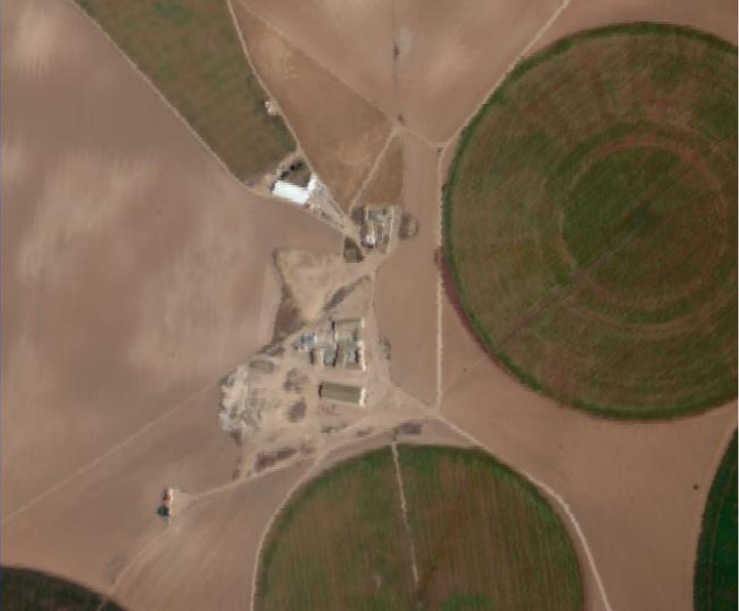}   
\IG[width=3.8cm]{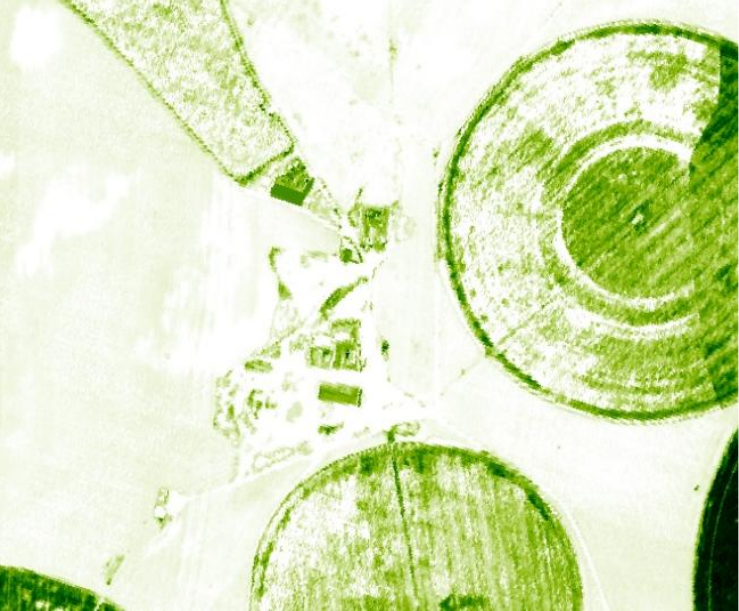} 
\IG[width=3.8cm]{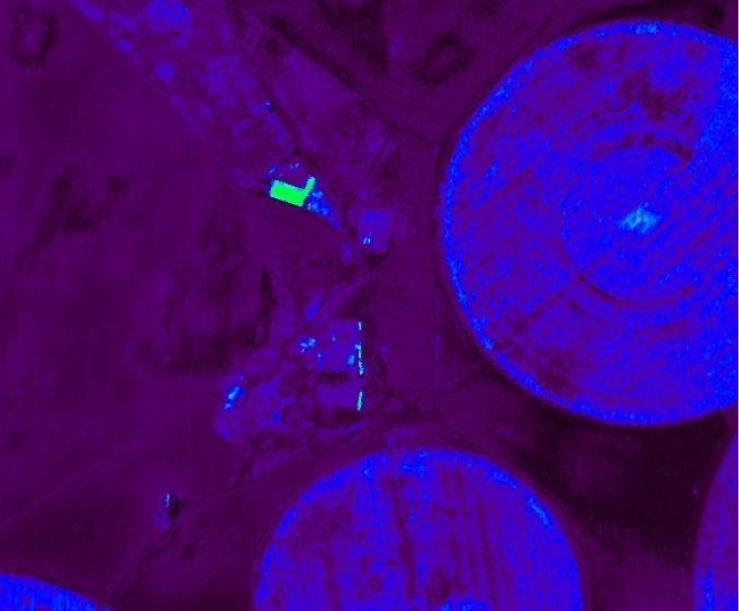}\\
\IG[width=3.8cm]{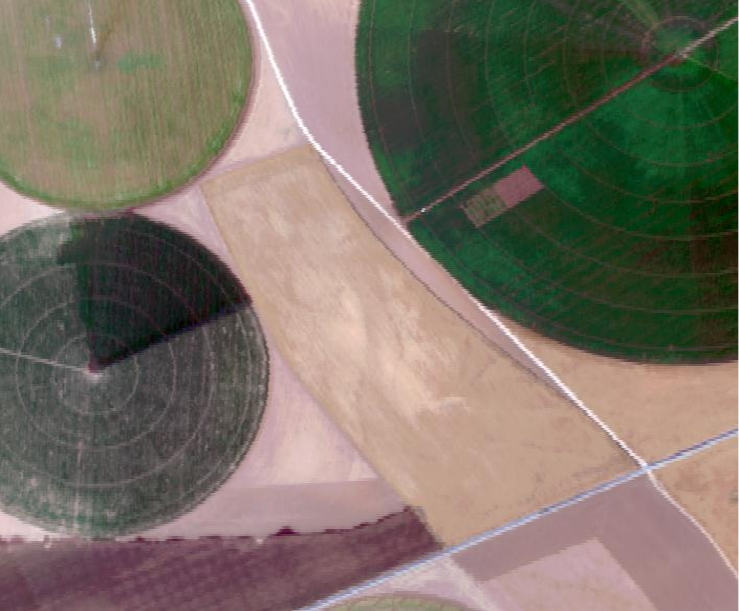}   
\IG[width=3.8cm]{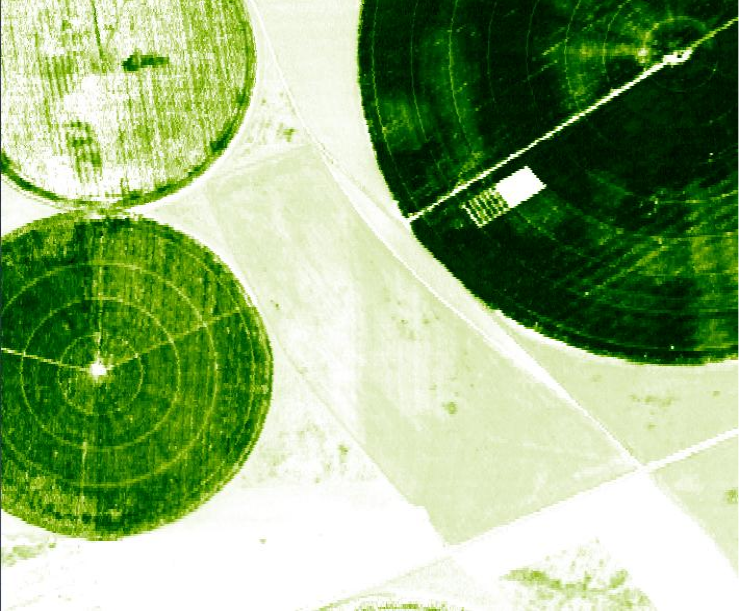} 
\IG[width=3.8cm]{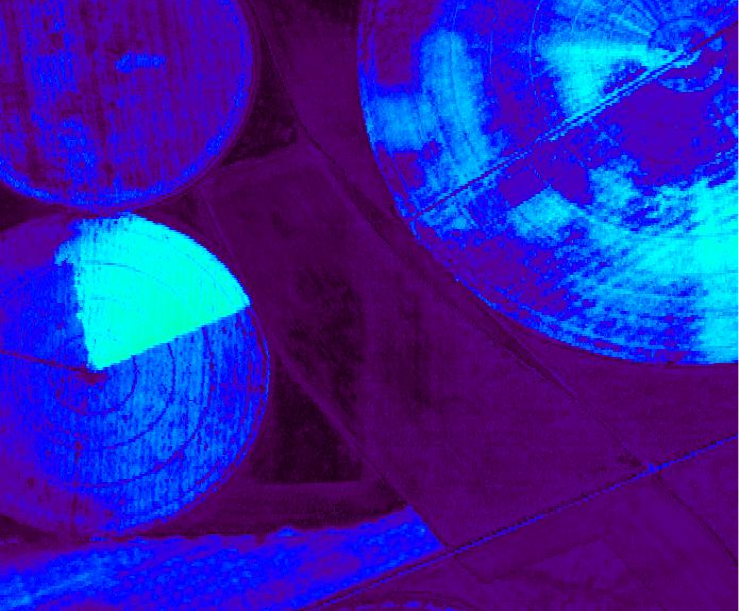}\\
\end{tabular}
\end{center}
\vspace{-0.25cm}
\caption{Three examples [top, middle, bottom] of CASI RGB snapshots [left], $Chl$ estimates [middle], and related uncertainty intervals [right].}  
\label{resultsCASIzoom2009}
\end{figure}

As a final example, uncertainty estimates were exploited to assess the robustness of the retrievals at multiple spatial scales. In~\cite{Verrelst2013a}, retrievals from hyperspectral airborne and spaceborne data over the Barrax area were compared. Based on the spareborne SPARC-2003 dataset, GP developed a model that was excellently validated ($r^{2}$: 0.96). The SPARC-trained GP model was subsequently applied to airborne CASI flightlines (Barrax, 2009) to generate $Chl$ maps. The accompanying uncertainty maps provided insight in the robustness of the retrievals. In general similar uncertainties were achieved by both sensors, which is encouraging for upscaling estimates from field to landscape scale. The high spatial resolution of CASI in combination with the uncertainties allows us to observe the spatial patterns of retrievals in more detail. Some examples of mean estimates and associated uncertainties are shown in Fig.~\ref{resultsCASIzoom2009}.

\section{Conclusions and further work}\label{sec:conclude}

This paper provided a comprehensive survey to the field of Gaussian Processes (GPs) in the context of remote sensing data analysis for Earth observation applications, and in particular for biophysical parameter estimation. We summarized the main properties of GPs and the advantages over other methods for estimation: essentially GPs can provide competitive predictive power, give error-bars for the estimations, allows to design and optimize sensible kernel functions, and also to analyze the encoded knowledge in the model via automatic relevance determination kernel functions.

GP models offer as well a solid Bayesian framework to formulate new algorithms well-suited to the signal characteristics. We have seen for example that by incorporating proper priors, we can encompass signal-dependent noise, and infer parametric forms of warping the observations as an alternative to either {\em ad hoc} filtering. On the downside, we need to mention the scalability issue: essentially, the optimization of GP models require computing determinants and invert matrices of size $n\times n$, which runs cubically in computational time and quadratically in memory storage. In the last years, however, great advances have appeared in machine learning and now it is possible to train GPs with several thousands of points. 

All the developments were illustrated at a local scales through a full set of illustrative examples in the field of geosciences and remote sensing. In particular, we treated important problems of ocean and land sciences: from accurate estimation of oceanic chlorophyll content and pigments, to vegetation properties from multi- and hyperspectral sensors.

\section*{Acknowledgments}
The authors wish to deeply acknowledge the collaboration, comments and fruitful discussions with many researchers during the last decade on GP models for remote sensing and geoscience applications: Miguel L\'azaro-Gredilla (Vicarious), Robert Jenssen (Univ. Troms\o, Norway), Martin Jung (MPI, Jena, Germany), and Sancho Salcedo-Saez (Univ. Alcal\'a, Madrid, Spain). 

This paper has been partially supported by the Spanish Ministry of Economy and Competitiveness under projects TIN2012-38102-C03-01 and ESP2013-48458-C4-1-P, and by the the European Research Council (ERC) consolidator grant entitled SEDAL with grant agreement 647423. AG is thankful to Marie Curie International Incoming Fellowship for supporting this work.
%%%%%%%%%%%%%%%%%%%%%%%%%%%%%%%%%%%%%%%%%%%%%%%%%%%%%%%%%%%%%%%%%%%%%%%%%%%
% REFERENCES

\end{document}

%% file: notation.tex
\newcommand{\Real}{\mathbb R}

\def\x{{\mathbf x}}

\newcommand{\vect}[1]{{\boldsymbol{\mathbf{#1}}}} % vector
\newcommand{\mat}[1]{{\boldsymbol{\mathbf{#1}}}} % matrix
\newcommand{\y}{{\vect y}}

\newcommand{\Kf}{{\mat K}_\vect{ff}}

\newcommand{\dataset}{{\cal D}}

\newcommand{\bigO}[0]{\mathcal{O}} 
\newcommand{\GP}[0]{\mathcal{GP}} 

\newcommand{\Normal}[0]{\mathcal{N}} 
 
\newcommand{\prob}{p}
 
\newcommand{\cut}[1]{} % cut out a part of the text